\documentclass[aps,pr,preprint,groupedaddress,
nofootinbib,longbibliography]{revtex4-1}

\usepackage{amsmath}
\usepackage{amssymb}
\usepackage{mathtools}
\usepackage{graphicx}

\usepackage[utf8]{inputenc}
\usepackage{lmodern}

\usepackage{mathrsfs}

\usepackage[hidelinks]{hyperref}

\usepackage{setspace}

\allowdisplaybreaks

\begin{document}

\title{Perturbative reduction of derivative order in EFT}

\author{Dra\v{z}en Glavan}
\email[]{drazen.glavan@fuw.edu.pl}
\affiliation{Institute of Theoretical Physics, Faculty of Physics,
University of Warsaw, Pasteura 5, 02-093 Warsaw, POLAND}

\date{\today}

\begin{abstract}

Higher derivative corrections are ubiquitous in effective field theories, which
seemingly introduces new degrees of freedom at successive order.
This is actually  an artefact of the implicit local derivative
expansion defining effective field theories. We argue that higher derivative 
corrections that introduce additional degrees of freedom should be removed
and their effects captured either by lower derivative corrections, or special 
combinations of higher derivative corrections not propagating extra derees of 
freedom. Three  methods adapted for this task are examined and field redefinitions
are found to be most appropriate. First order higher derivative corrections 
in a scalar tensor theory are removed by field redefinition and it is found that
their effects are captured by a subset of Horndeski theories. A case is made for
restricting the effective field theory expansions in principle to only terms not 
introducing additional degrees of freedom.

\end{abstract}

\pacs{FIND PACS}

\maketitle



\section{Introduction}
\label{sec: Introduction}

Effective theories are a physical framework based on an observation that 
not all scales of nature are relevant for describing of a particular given 
system~\cite{Wells:2012rla,Burgess:2007pt,Donoghue:1995cz}.
Usually it pertains to descriptions of some low energy phenomenon
in terms of the degrees of freedom and interactions relevant for that system,
with typical energy scales well below some higher scales typical for the interactions
and degrees of freedom which we neglect. It is not to say that we have
to neglect the effect of physics on higher scales entirely. We may study
the small corrections that it supplies to our low energy system,
but we do not need to know exactly what that physics is. Its effects 
can be captured in terms of the degrees of freedom of the low energy system
by introducing additional effective interactions suppressed
by the ratio of scales typical for the system at hand and the high scale
of new physics. In field theories such effects are encoded
by all the possible corrections to the leading order action organized
as an expansion in the inverse powers of the high energy (or mass) scale.
At each order all that is demanded from the correction terms is 
to satisfy the assumed symmetries and to have the right dimensionality.

Higher derivative corrections in effective field theory (EFT) are 
ubiquitous, and are very hard to avoid. Such corrections, taken at face value,
tend to exhibit unphysical behaviour. The issues regarding them
were most recently discussed by Burgess and Williams~\cite{Burgess:2014lwa},
who also argued we typically need not worry about the said issues.
Here we make a case for the issues actually not being there if one
treats higher derivatives in EFTs correctly in the same perturbative 
spirit in which they arise in the first place\footnote{During the preparation
of this paper, Solomon and Trodden have published a 
preprint~\cite{Solomon:2017nlh} discussing the higher derivative issues in EFTs 
from  the same perspective as we do here.}

It is known that higher derivative theories in general exhibit
unphysical behavior.
The most well known issue encountered is that 
of Ostrogradsky instabilities~\cite{Ostrogradsky:1850fid,Woodard:2015zca}.
Higher derivative theories generally contain more degrees of freedom per dynamical 
variable as compared to the lower derivative theories, and such theories
we refer to as {\it genuine} higher derivative theories in this paper. Usually these extra 
degrees of freedom are ghost-like, and their excitation can lower the energy
of the system without bounds. This is always true for theories without constraints,
but can in certain cases be avoided in constrained theories (e.g. $f(R)$ gravity).
Another issue often exhibited by higher derivative theories is the 
existence of runaway solutions. It is important to emphasize that there are
theories with higher derivatives in the action that are not genuine in the
sense defined above, since they contain the same number of degrees of
freedom as their lower order counterparts with the same number of dynamical
variables. Examples of such non-genuine higher derivative theories
are the Horndeski theory~\cite{Horndeski:1974wa},
generalized Galileons~\cite{Nicolis:2008in,Deffayet:2009mn}, and more generally
the beyond Horndeski theories~\cite{Gleyzes:2014dya,Gleyzes:2014qga}, and all the  the DHOST theories~\cite{Langlois:2015cwa,Crisostomi:2016czh,BenAchour:2016fzp}, 
which are defined as the most general scalar-tensor
theories with three propagating degrees of freedom.
These are studied very much today in the context of the effective field
theory of dark energy~\cite{Gubitosi:2012hu,Tsujikawa:2014mba}
For convenience, we sometimes refer to non-genuine higher derivative theories 
as lower derivative theories when it is obvious we are emphasizing their
property of not carrying extra degrees of freedom.

Effective field theories in actuality are not genuine higher derivative theories.
This becomes clear when we examine cases where we can actually derive 
them from a more fundamental theory by integrating out
some heavy degrees of freedom. 
In principle, we do not need to assume much about the degrees of freedom
being integrated out. This procedure produces an effective action for the 
system which contains nonlocal terms which do not carry extra 
degrees of freedom. By construction, the procedure
of integrating out degrees of freedom can only lower their total number
in the resulting theory. Only when we decide that the degrees of freedom
integrated out are heavy with respect to some energy scale we are interested
in can the nonlocal terms in the action be considered small, and expanded 
in inverse powers of the heavy mass scale. This expansion corresponds
to the local derivative expansion of the nonlocal terms, and this is
how higher derivative terms appear in the low energy EFT. The modern
approach to field theories is to consider them as EFTs up to some
higher energy scale. Then naturally all field theories can be considered to 
have corrections suppressed by this high energy (mass) scale. But here
we maintain that these corrections ought to descend from some in principle
non-local corrections.

At successively higher orders in EFTs terms with higher and higher 
derivatives appear. It makes no physical sense that each order of perturbation
theory introduces new degrees of freedom. Rather, this is an artifact of
our approximation scheme. 
The observation that
the higher derivative terms appear perturbatively, and that their effects should
be taken into account perturbatively as well,
resolves the issue of extra degrees of freedom. They never appear in
perturbatively expandable solutions.

This problem was first encountered and studied in the context of
Abraham-Lorentz theory of the electron~\cite{Jackson:1998nia}, where the
third time derivative of the electron's position appears in the equation
of motion. This third derivative is there to take into account the 
electron's reaction to its own radiation. The space of solutions of this theory
contains some clearly unphysical runaway solutions -- one can
set initial conditions for the electron to be at rest, and it can still
accelerate to velocities arbitrarily close to the speed of light. This is an example
of the runaway solution due to equations of motion requiring the specification
of the initial acceleration in addition to the initial position and velocity.
It is an unphysical behavior introduced by treating the higher
derivative corrections on the same footing as the leading order equation.
One must discard all the solutions of the equations of motion that are
not analytic in the expansion parameter as unphysical.

The aim of this paper is to investigate three different methods of
correctly accounting for the higher derivative terms in the EFT expansion.
They all aim to reduce the derivative order of the higher derivative corrections,
representing to them in terms on lower derivative ones,
and hence we refer them as {\it derivative reduction} methods. In particular
we consider (i) derivative reduction on the level of the equations of motion,
(ii) derivative reduction by method of perturbative constraints (described in 
Sec.~\ref{sec: Formalism}), and (iii) derivative reduction via field redefinition.
Their advantages and disadvantages, utility and drawbacks are analyzed 
in detail in Sec.~\ref{sec: Scalar} on a series of examples. They all serve 
to motivate the overall claim we wish to make about the nature of the EFT expansion 
in general: {\it Genuine higher derivative corrections appearing in EFTs are completely 
captured by the most general  corrections not introducing
extra degrees of freedom}.

It is common lore, mostly due to the work of Simon and 
Parker~\cite{Simon:1990ic,Parker:1993dk}, that
the higher derivative terms can always be removed from the equations
of motion by perturbatively applying lower order equations of motion.
This is certainly true for any single particle systems, such as the
minisuperspace model they considered, where the only dynamical variable
is the scale factor (see also~\cite{Mazzitelli:1991rf}). Even though the study of this 
single particle models and insightful, 
multiple issues and subtleties of this derivative reduction method are missed.
One of the 
things that we examine in this paper is the analogous derivative 
reduction procedure at the equation of motion level in the
field theory system possessing Lorentz symmetries, and whether this
derivative reduction procedure is consistent with maintaining Lorentz symmetry. 
Our findings are that in general it is not, without allowing
for field redefinitions. The same conclusion ought to be valid when
considering local symmetries, such as in the EFT of gravity.

Reduction of the derivative order at the level of the equations of motion
(if needed supplemented by field redefinition) is sufficient if one is
interested in obtaining solutions as power series in the expansion
parameter truncated to a given order (for some applications of this strategy to 
EFT of gravity see~\cite{Cremonini:2009ih,Goon:2016mil}). 
Sometimes such solutions are not sufficient. 
This is often the case in nonequilibrium systems~\cite{Berges:2004yj},
where spurious secular phenomena might be introduced by considering 
perturbative corrections to the leading order solutions. In such cases
a resummation scheme is called for, which, in broad strokes, corresponds
to solving exactly the truncated EFT equation with reduced derivatives.
For instance, such circumstances arise when studying cosmological
solutions in general relativity with EFT corrections.
The problem that arises here is that the derivative reduction at the level of
the equations of motion is not unique, in the sense that multiple
equations accomplish the same task of capturing perturbative solutions
to a given order. Attempting the resummed solution for different
equations might lead to rather different results, and even exhibit effects
such as energy non-conservation in conservative 
systems\footnote{{\setstretch{1.0}This point is understood in the context of the 2PI
resummation scheme in QFT. By consistently truncating the Dyson-Schwinger 
equations it is possible to derive a family of equations that are
all equivalent to the 2PI equation to the given order, but only one of them
corresponds to the 2PI one which can be derived from an action principle. 
It is said that the other equations {\it overcount} the diagrams.}} The correct way
to avoid such spurious effects is to base the resummed solutions on 
the equations of motion that derive from an action principle, which
guarantees properties such as energy conservation to all orders.
The consideration of nonequilibrium systems motivates us to
develop a formalism for reducing the derivative order of EFT
corrections directly on the level of the action. Most of this work is devoted
to the investigation of such methods.

\vskip+0.5cm

Higher derivative EFT corrections are particularly awkward if one wishes
to consider them in the canonical (Hamiltonian) formalism. The initial 
motivation for this work
came from the problem of constructing and quantizing 
higher order gauge-invariant 
cosmological perturbations\footnote{Higher order
here refers to higher order in infinitesimal diffeomorphism transformations \\} 
in a single scalar inflationary 
model~\cite{Prokopec:2012ug,Prokopec:2013zya,Domenech:2017ems},
with the idea of addressing certain issues regarding the loop correction
to the primordial power spectrum (see~\cite{Miao:2012xc} and references therein).
The formalism most suited for
finding the gauge-invariant perturbations seems to be the canonical one,
since it makes the constraint structure of the theory explicit~\cite{Langlois:1994ec}.
The quantization of these higher order perturbations necessarily requires
the introduction of higher derivative terms to the action in order
to guarantee one-loop renormalizability~\cite{Birrell:1982ix}. But these
higher derivative terms alter the structure of cosmological perturbations
beyond linear order, and should be considered right from the start. Hence
we need to consider these higher derivative terms in the canonical formalism.

In the canonical formalism every degree of freedom is promoted to an
independent field, including the degrees of freedom associated to 
higher derivative terms in the EFT expansion~\cite{Gitman:1990qh}.
In the EFTs this is very cumbersome, since the resulting Hamiltonian
is no longer perturbative in the inverse powers of the heavy mass scale 
assumed to be the control
parameter in the EFT expansion, but rather has term containing
positive powers of this mass parameter. This is clearly a signal of
doing something wrong. In fact, it means precisely what we have
stated, considering higher derivative terms to introduce extra degrees of freedom
is inconsistent with the suppositions of EFT. 

The very derivation of the canonical formalism in the presence of higher
derivative EFT corrections has to be modified to accommodate for the
requirement of perturbativity in the expansion parameter. This was 
noted and clarified by Jaen, Llosa, and Molina,~\cite{Jaen:1986iz}, and by
Eliezer and Woodard~\cite{Eliezer:1989cr}. The perturbativity requirement
effectively introduces perturbative second-class constraints into the 
theory~\cite{dirac2001lectures},
which when solved for reduce the total number of degrees of freedom 
to the correct one, resulting in the lower order phase space action.
Section~\ref{sec: Formalism}
is devoted to recasting this method in a somewhat different form,
and to the discussion of its possible drawbacks. Such a method would be 
preferable as a way to reconcile higher derivative corrections with the
physical assumptions EFT expansion makes, since it would
automatically mean that perturbativity requirement maps
 higher derivative EFT corrections onto
lower derivative EFT ones. Larger part of Sec.~\ref{sec: Scalar} is
devoted to applying this method to a scalar field system, where
it is shown that even though the method always accomplishes what
it is designed to do -- produces a lower derivative phase space action --
it in general does not provide a configuration space action (the
Lagrangian) which makes it rather impractical for applications.

\vskip+0.5cm

The third method of derivative reduction that we consider in this work
is the field redefinition method. It ought be true that the physics
(whatever is measured in the experiments)
should not be altered by one's choice of field variables describing
the system. Hence, it is in principle permissible to perform transformations
from one set of field variables to another (field redefinitions).
There is a physical requirement on these redefinitions
 that they should not change the number of degrees of freedom
 system. Considering derivatives of the field, it means they should not 
 contain any. In other words, we are not allowed to make a higher 
 derivative theory out of the lower derivative one. 
 
Consider now field redefinitions is the EFTs. Assume that we know
 the exact form of the nonlocal correction to the action, before it is approximated
by a local derivative expansion. Now we can imagine making a nonlocal
field redefinition in such an action, resulting in another nonlocal action.
We can take  this  field redefinition to be  a special one 
-- by requiring that the resulting nonlocal action has a local derivative expansion
that contains no genuine higher derivative terms. Such special nonlocal
transformation thus eliminates spurious degrees of freedom from
the low energy EFT.  Now consider taking
a short-cut of this construction, directly on the low energy EFT. It is
equivalent to making a local field redefinition containing derivatives,
that removes all the genuine higher derivative terms from the EFT
expansion (to a given order). It is due to these considerations 
that field redefinitions containing derivatives are allowed in EFTs,
in the sense that field redefinitions themselves are treated as an 
EFT local derivative expansion. It is found in this paper that the
field redefinition method of the derivative reduction is in general 
superior to the remaining two, especially when considering symmetries
of the field systems.

\vskip+0.5cm

The three mentioned methods for reducing the derivative order of EFT
corrections are studied in this paper on several examples of a single scalar
field in flat space. Attention is
paid to the compatibility of these methods with Lorentz symmetry, and the 
possibility of obtaining a configuration space (Lagrangian) formulation of the
reduced system. Field redefinitions are found to be the preferable method.
The field redefinition method is then applied to the first order EFT
corrections of a scalar-tensor theory, where it is found that the corrections are
captured by the subset of Horndeski actions, which are the most general 
scalar-theories possessing second order equations of motion.

 \vskip+0.5cm

The introductory section that is drawing to a close serves to define and motivate 
the study of higher derivatives in the context of effective field theories. 
Section~\ref{sec: Formalism} is devoted to the presentation of the method of
derivative reduction via perturbative constraints. In Section~\ref{sec: Scalar} we 
apply the three derivative reduction methods to four cases of a single scalar
field with higher derivative corrections. The lessons learned from these 
examples are summarized at the end of Sec.~\ref{sec: Scalar}. In Section~\ref{sec: ST}
the field redefinition method is employed to eliminate the first order higher
derivative corrections from the EFT expansion of a scalar-tensor theory.
The last section contains the discussion of the results and the concluding remarks.


\section{Method of perturbative constraints}
\label{sec: Formalism}

This section presents the method of perturbative derivative reduction
by implementing perturbative constraints\footnote{Dubbed {\it localization}
by the authors of~\cite{Eliezer:1989cr}.}.
The method has been developed for systems without
constraints  by Ja\'{e}n, Molina and Llosa~\cite{Jaen:1986iz}, 
and by Eliezer and Woodard~\cite{Eliezer:1989cr}. Here we
present a modified version of it, which has a few advantages:
(i) it applies to systems with arbitrary leading order
actions, not only to the case of a free leading order system, and 
(ii) it applies to explicitly time dependent systems, and 
ones where Noether current is cumbersome to define,
(iii) it is easily generalized to systems with constraints.

Even though in this work we aim to investigate field systems,
the method here is presented for a multi-particle systems.
The reason is that the notation is simpler, and we do not have to
explicitly write gradients, which just obscures the expressions. 
Nevertheless, the generalization to field systems is obvious and 
straightforward.

Throughout this section, and later when applying the method we use 
Dirac's notation for equalities~\cite{dirac2001lectures}, where we distinguish
between strong equality (off-shell equality) ``$=$'' denotes the equality
which is true on the level of the action, without applying the equations
of motions, and a weak equality (on-shell equality) ``$\approx$''
designating equalities valid only after the imposition of equations of motion.

Say that we are interested in studying a system of particles described by
variables $x_i$, that at leading order is specified by the {\it leading order action},
\begin{equation}
S^{(0)}[x_i] = \int\! dt \, \mathscr{L}^{(0)}(x_i, \dot{x}_i) \, ,
\end{equation}
which is assumed not to be singular, {\it i.e.} its Hessian is supposed invertible,
\begin{equation}
{\rm det} \Bigl( \frac{\partial^2 \mathscr{L}^{(0)}}{\partial \dot{x}_i \partial\dot{x}_j} 
	\Bigr) \neq 0 \, .
\end{equation}
and there are no constraints. The generalization of this method to systems
with constraints is beyond the scope of this paper.
We consider the first $N$ corrections to this action in the form of an EFT 
expansion in terms (operators) of successively higher dimension,
with $\epsilon$ being the expansion parameter, given by the configuration
space action,
\begin{equation}
S[x_i] = S^{(0)}[x_i] + \epsilon S^{(1)}[x_i] + \epsilon^2S^{(2)}[x_i]
	+ \dots + \epsilon^N S^{(N)}[x_i]
	= \sum_{n=0}^{N} \epsilon^n S^{(n)}[x_i] \, .
\label{CSA}
\end{equation}
The higher order corrections are also assumed to contain 
irreducible higher time derivatives\footnote{Higher time derivatives that 
can be removed by partial integration are considered reducible.},
with each order introducing one more derivative,
\begin{equation}
S^{(n)}[x_i] = \int\! dt \,
	\mathscr{L}^{(n)} \bigl( x_i, \dot{x}_i, \dots, x_i^{[n]}, x_i^{[n+1]} \bigr) \, ,
\qquad
x_i^{[m]} \equiv \frac{d^m}{dt^m} x_i \, ,
\end{equation}
so that each new order introduces one derivative more.
In what follows we construct a method to isolate a subspace
of solutions of this theory that are regular in the expansion parameter
$\epsilon$, {\it i.e.} that are perturbatively expandable in $\epsilon$.

The starting point of the method is to define the 
{\it extended action} ~\cite{Gitman:1990qh}, which can be seen
as the intermediate step between the Lagrangian and the
Hamiltonian formalism, and is particularly suitable for theories
with constraints.
The extended action is defined by promoting all the time 
derivatives to independent velocity variables,
$x_i^{[n]} \rightarrow V_{i}^n$, and introducing Lagrange 
multipliers $\pi_i$, $\pi_i^1$,\dots,$\pi_i^{N-1}$ to ensure
the on-shell equivalence of the two formulations,
\begin{align}
\MoveEqLeft[4]
\mathcal{S}[x_i, V^1_i, V^2_i, \dots, V^{N+1}_i, \pi_i,\pi_i^1,\dots,\pi_i^N] 
\nonumber \\
={}& \int\! dt \, \biggl\{
	\mathscr{L}^{(0)}(x_i, V^1_i)
	+ \epsilon \mathscr{L}^{(1)}(x_i, V^i_i, V^2_i)
	+ \dots	
	+ \epsilon^N \mathscr{L}^{(N)}(x_i, V^1_i, \dots, V^{N+1}_i)
\nonumber \\
&	\hspace{2.5cm}
	+ \pi_i (\dot{x}_i - V^1_i) + \pi_i^1 (\dot{V}_1-V^2_i) + \dots
	+ \pi_i^{N} (\dot{V}^{N}_i - V^{N+1}_i)
	\biggr\}
\label{EA}
\end{align}
The canonical formalism, namely the phase space action $\mathscr{S}$ 
is now one step away. One needs to solve
for the highest velocities $V_i^{N+1}$ from the equation descending from
the variation of the extended action with respect to $V_i^{N+1}$'s,
\begin{equation}
\frac{\delta \mathcal{S}}{\delta V^{N+1}_i}
	= \epsilon^N \frac{\partial \mathscr{L}^{(N)}}{\partial V^{N+1}_i}
	- \pi_i^{N} \approx 0 \, .
\label{V^N equation}
\end{equation}
which is an algebraic relation involving no time derivatives
and plug it back into $\mathcal{S}$. The solutions for $V_i^N$'s,
\begin{equation}
V^{N+1}_i \approx \overline{V}^{N+1}_i(x_i, V^1_i,\dots V^{N}_i, 
	\pi_i,\pi^1_i, \dots \pi^{N}_i ) \, ,
\end{equation}
are then plugged back into $\mathcal{S}$ as strong equalities,
which results in the phase space action,
\begin{align}
\MoveEqLeft[4]
\mathscr{S}[x_i,\pi_i, V^1_i, \pi^1_i, \dots, V^{N}_i, \pi^{N}_i]
\nonumber \\
={}& \int\! dt \, \biggl\{ \pi_i \dot{x}_i + \pi^1_i \dot{V}^1_i
	+ \dots + \pi^N_i \dot{V}^N_i 
	- H\bigl(x_i,\pi_i, V^1_i, \pi^1_i, \dots V^{N}_i, \pi^{N}_i \bigr) \biggr\} \, ,
\end{align}
where the Hamiltonian is
\begin{align}
H ={}& \pi_i V^1_i + \pi^1_i V^2_i + \dots 
	+ \pi^{N-1}_i V_i^N 
	+ \pi^{N} \overline{V}_i^{N+1}
\nonumber \\
&	- \mathscr{L}^{(0)}(x_i, V^1_i) - \epsilon \mathscr{L}^{(1)}(x_i, V^1_i, V^2_i)
	- \dots - \epsilon^N \mathscr{L}^{(N)}(x_i, V^1_i, V^2_i,\dots, \overline{V}^{N+1}_i) \, .
\end{align}
This phase space action is on-shell equivalent to the extended 
action~(\ref{EA}),
and hence to the starting configuration space action~(\ref{CSA}) as well.
This construction is equivalent 
to the one introduced by Ostrogradsky~\cite{Ostrogradsky:1850fid,Woodard:2015zca}.
There is nothing about this construction specific to higher derivative theories.
It is perfectly applicable to standard lower derivative theories, and 
in fact allows for a much more pedagogical introduction of the canonical
(Hamiltonian) formalism.

The obstruction to the construction of the phase space action in the
case at hand is that equation~(\ref{V^N equation}) does not admit 
solutions for $V_i^{N+1}$'s analytic in the perturbative expansion 
parameter $\epsilon$, and is therefore not admissible by our 
assumptions\footnote{As the authors of~\cite{Jaen:1986iz} 
put it, the variables we are dealing
with are a quotient ring $\mathbb{R}/\epsilon^{N+1}$, {\it i.e.} 
polynomials in $\epsilon$ with real coefficients, modulo $\epsilon^{N+1}$.}.
Given such assumptions, the equation~(\ref{V^N equation}) 
actually represents a constraint,
and should be treated as such. This observation 
allows us to isolate from the full theory only the subset of solutions
perturbatively expandable in $\epsilon$. 

In the language of Dirac's
canonical formalism~\cite{dirac2001lectures} equation~(\ref{V^N equation})
is a {\it primary} constraint. The requirement that this constraint is consistent,
{\it i.e.} that its time derivative vanishes on-shell produces a 
{\it secondary} constraint. Again requiring consistency, and repeating this process
generates a set of constraints that actually end up being {\it second-class},
meaning that in the end the variables $V_i^{N+1}$'s are uniquely determined
(by the equations of motion).
Even though we can cast the whole method in the form usually 
employed in canonical formalism, by defining Poisson brackets, and
working with the Hamiltonian, it is actually not the most practical way to
proceed here. It is more convenient to work with the extended action~(\ref{EA}),
and the equations of motion descending from varying it with respect to
independent variables.

The systematic way to proceed is to write down all the equations 
descending from varying the extended action,
\begin{eqnarray}
&\displaystyle \frac{\delta\mathcal{S}}{\delta V_i^{N+1}} \approx 0
\qquad &\Rightarrow \qquad 
\pi_i^{N} \approx \epsilon^N \frac{\partial \mathscr{L}^{(N)}}{\partial V_i^{N+1}} \, ,
\label{first eq}
\\
&\displaystyle \frac{\delta\mathcal{S}}{\delta V_i^{N}} \approx 0
\qquad & \Rightarrow \qquad 
\dot{\pi}_i^{N} \approx - \pi_i^{N-1} 
	+ \epsilon^N \frac{\partial \mathscr{L}^{(N)}}{\partial V_i^{N}}
	+  \epsilon^{N-1} \frac{\partial \mathscr{L}^{(N-1)}}{\partial V_i^{N}} \, ,
\\
&\displaystyle \vdots \qquad& \nonumber 
\\
&\displaystyle \frac{\delta\mathcal{S}}{\delta V_i^2} \approx 0
\qquad & \Rightarrow \qquad 
\dot{\pi}_i^{2} \approx - \pi_i^1
	+ \epsilon^N \frac{\partial \mathscr{L}^{(N)}}{\partial V_i^{2}}
	+ \dots + \epsilon^{1} \frac{\partial \mathscr{L}^{(1)}}{\partial V_i^{2}}  \, ,
\\
&\displaystyle \frac{\delta\mathcal{S}}{\delta V_i^1} \approx 0
\qquad & \Rightarrow \qquad 
\dot{\pi}_i^1 \approx - \pi_i
	+ \epsilon^N \frac{\partial \mathscr{L}^{(N)}}{\partial V_i^{1}}
	+ \dots 
	+ \epsilon^{0} \frac{\partial \mathscr{L}^{(0)}}{\partial V_i^{1}}   \, , \qquad
\\
&\displaystyle \frac{\delta\mathcal{S}}{\delta x_i} \approx 0
\qquad & \Rightarrow \qquad 
\dot{\pi}_i \approx \epsilon^N \frac{\partial \mathscr{L}^{(N)}}{\partial x_i}
	+ \dots 
	+ \epsilon^{0} \frac{\partial \mathscr{L}^{(0)}}{\partial x_i}   \, , \qquad
\\
&\displaystyle \frac{\delta\mathcal{S}}{\delta \pi_i} \approx 0
\qquad &\Rightarrow \qquad 
\dot{x}_i \approx V_i^1 \, ,
\\
&\displaystyle \frac{\delta\mathcal{S}}{\delta \pi_i^1} \approx 0
\qquad &\Rightarrow \qquad 
\dot{V}_i^1 \approx V_i^2 \, ,
\\
&\displaystyle \vdots \qquad& \nonumber
\\
&\displaystyle \frac{\delta\mathcal{S}}{\delta \pi_i^{N}} \approx 0
\qquad &\Rightarrow \qquad 
\dot{V}_i^{N} \approx V_i^{N+1} \, .
\label{last eq}
\end{eqnarray}
These equations need to be satisfied at each order in $\epsilon$.
Therefore, we may solve them order by order. Multiplying all of the equations
by $\epsilon^N$ yields the leading order equations,
\begin{align}
\epsilon^N \pi_i^N \approx{}& 0 \, ,
\label{first}
\\
\epsilon^N \dot{\pi}_i^N \approx{}& - \epsilon^N \, ,
\\
\vdots \ \ \nonumber 
\\
\epsilon^N \dot{\pi}_i^1 \approx{}& - \epsilon^N \pi_i \, ,
\\
\epsilon^N \dot{\pi}_i \approx{}& 
	\epsilon^N \frac{\partial \mathscr{L}^{(0)}}{\partial x_i} \, ,
\\
\epsilon^N \dot{x}_i \approx{}& \epsilon^N V_i^N \, ,
\\
\vdots \ \ \nonumber 
\\
\epsilon^N \dot{V}_i^N \approx{}& \epsilon^N V_i^{N+1} \, .
\end{align}
The first equation~(\ref{first}) is a constraint. Taking derivative of it 
generates another constraint. Taking derivative of that one generates 
yet another one, etc. The result of this process are leading order solutions
of the constraints,
\begin{equation}
\epsilon^N \pi_i^n \approx \epsilon^N \overline{\pi}_i^{n}(x_i, \pi_i) \, ,
\qquad
\epsilon^N V_i^n \approx \epsilon^N \overline{V}_i^{n}(x_i,\pi_i) \, ,
\end{equation}
where we have determined just the leading order 
term in the $\epsilon$ expansion 
of~$\overline{\pi}_i^{n}(x_i, \pi_i)$ and~$\overline{V}_i^{n}(x_i,\pi_i)$.

Next step is to multiply equations~(\ref{first eq}-\ref{last eq}) 
by $\epsilon^{N-1}$ and repeat the same process to obtain
\begin{equation}
\epsilon^{N-1} \pi_i^n \approx \epsilon^{N-1} \overline{\pi}_i^{n}(x_i, \pi_i) \, ,
\qquad
\epsilon^{N-1} V_i^n \approx \epsilon^{N-1} \overline{V}_i^{n}(x_i,\pi_i) \, ,
\end{equation}
meaning that we now have first two terms in the $\epsilon$ expansion
of~$\overline{\pi}_i^{n}(x_i, \pi_i)$ and~$\overline{V}_i^{n}(x_i,\pi_i)$.

 This is repeated until 
we have {\it solved for all the constraints} to obtain
\begin{equation}
\pi_i^n \approx \overline{\pi}_i^{n}(x_i, \pi_i) \, ,
\qquad
V_i^n \approx \overline{V}_i^{n}(x_i,\pi_i) \, ,
\end{equation}
to order $\epsilon ^N$. The final step involves plugging these of-shell
relations as strong equalities into the extended action~(\ref{EA}).
This produces the reduced phase space action,
\begin{equation}
\mathscr{S}_{\rm red}[x_i,\pi_i] = \int\! d^{D\!}x \, \biggl\{
	\pi_i \dot{x}_1 + \overline{\pi}_i^1 \dot{\overline{V}}_i^1 + \dots
	+ \overline{\pi}_1^N \dot{\overline{V}}_i^N - H_{\rm red}(x_i,\pi_i)
	\biggr\} \, ,
\label{reduced phase action}
\end{equation}
where the reduced Hamiltonian is
\begin{align}
H_{\rm red}(x_i,\pi_i) ={}& \overline{\pi}_i \overline{V}_i^1
	+  \overline{\pi}_i^1 \overline{V}_i^1 + \dots 
	+  \overline{\pi}_i^N \overline{V}_i^{N+1}
	- \mathscr{L}^{(0)}(x_i, \overline{V}_i^1) 
\nonumber \\
&	
	- \epsilon \mathscr{L}^{(1)}(x_i, \overline{V}_i^1, \overline{V}_i^2)
	- \dots
	- \epsilon^N \mathscr{L}^{(N)}
		(x_i, \overline{V}_i^1, \overline{V}_i^2,\dots, V_i^N) \, .
\end{align}
This phase space action now depends explicitly only on the
degrees of freedom found in the leading order action.

The nonstandard feature of the phase space action~(\ref{reduced phase action})
is the non-canonical kinetic term (the part containing the time derivatives).
The reader familiar with Dirac's canonical formalism will recognize this
actually defining the Dirac brackets between variables. We are not 
interested in the brackets here so we will not work them out
(for examples of this see~\cite{Eliezer:1989cr}). We are interested though
whether there is a configutation space equivalent of the reduced phase space
action~(\ref{reduced phase action}). The symplectic part plays a crucial 
role in the answer to this question. 

Since $\overline{V}_i^n$'s and $\overline{\pi}_i^n$'s are functions of 
$x_i$ and $\pi_i$ only, the symplectic part of the action can be written as
\begin{equation}
\int\! d^{D\!}x \, \Bigl\{
	\bigl[ \pi_i + \epsilon\Pi_i( x_j, \pi_j) \bigr] \dot{x}_i 
	+ \epsilon X_i (x_j, \pi_j) \dot{\pi}_i
	\Bigr\} \, ,
\end{equation}
where $\Pi_i$ and $X_i$ have some expansion in $\epsilon$.
In general no amount of partial integrations can remove 
the time derivative from $\pi_i$ completely. In such a case
one cannot follow the standard procedure to obtain the 
configuration space action. Usually what one does
when there is no $\dot{\pi}_i$ is 
solve for the conjugate momenta from the Hamilton's equations
as functions of $x_i$ and $\dot{x}_i$,
\begin{equation}
\pi_i \approx \overline{\pi}_i (x_i,\dot{x}_i) \, ,
\end{equation}
and then plugs this into the phase space action as a strong equality.
This then results in the configuration space action, and the Lagrangian
formulation. The reason why it works is that it can easily be shown 
that two formulations are on-shell equivalent. But when $\dot{\pi}_i$
is present in the symplectic part of the action in an irreducible way
this procedure does not result in the on-shell equivalent configuration
space action, meaning there is no Lagrangian formulation of the reduced
system only in terms of $x_i$'s. Note that this never happens for
systems with a single degree of freedom where $\dot{\pi}$ can always
be removed\footnote{This should not to be confused with systems with
a single {\it local} degree of freedom, such as the scalar field, 
which actually has uncountably many degrees of freedom.}.

One can remove the irreducible $\dot{\pi}_i$'s from the symplectic part
by a redefinition of the variables $x_i$'s. Unfortunately this procedure
is not unique, and more likely than not one will make a redefinition
which for example breaks manifest Lorentz covariance. Even though
there is nothing fundamentally wrong with that, this is reason why we
shy away from making variable redefinitions in the phase space action,
since it is very likely to complicate things. That being said, if one is 
interested only in the phase space formulation, 
action~(\ref{reduced phase action}) can always be brought to the canonical
form by redefining the variables and their conjugate momenta 
perturbatively in $\epsilon$.

The method presented in this section is applied to a few cases involving 
a single scalar field in the next section. If the reader find the presentation
of the method in this section too dense, two examples in the next section 
are worked out in detail.


\section{Derivative reduction in single scalar EFT }
\label{sec: Scalar}

This section serves to test, examine, and compare the three methods
of perturbative derivative reduction on simple examples of real
scalar field systems. Note that this is not a one-particle system,
but a system with one {\it local} degree of freedom, interacting 
with the neighboring ones via gradients. We work in $D$ spacetime
dimensions, with a positive signature Minkowski metric; the
D'Alembertian is
$\square=\partial_\mu \partial^\mu$, and the Laplacian $\partial_i \partial^i$,
and a dot denotes a time derivative, while a prime denote a derivative with
respect to the field.

We consider an EFT expansion in the system of a single scalar field,
up to cubic order in the expansion parameter $\epsilon=M^{-2}$,
where $M$ is some heavy mass scale,
\begin{equation}
S[\phi] = S^{(0)} [\phi] + \epsilon S^{(1)}[\phi] + \epsilon^2 S^{(2)}[\phi]
	+ \epsilon^3 S^{(3)}[\phi] + \mathcal{O}(\epsilon^4) \, ,
\end{equation}
where the leading order action is taken to be a scalar with a canonical 
kinetic term, and a potential term,
\begin{equation}
S^{(0)} = \int\! d^{D\!}x \, \biggl\{
	- \frac{1}{2} (\partial_\mu\phi) (\partial^\mu\phi) - U(\phi) \biggr\} \, .
\label{leading order action}
\end{equation}
The corrections to this action are assumed to be all possible
Lorentz invariant  terms of certain mass dimension, and
the expansion is ordered in the number of these mass dimensions.
The first order correction,
\begin{equation}
\epsilon S^{(1)} = \epsilon \sum_{n=1}^{3} \alpha_n^{(1)} S^{(1)}_n \, ,
\label{first order action}
\end{equation}
is composed out of all possible independent
dimension 6 terms (operators),
\begin{align}
S^{(1)}_1 ={}& \int\! d^{D\!}x \, \phi^6 \, ,
\label{Ss(1)1}
\\
S^{(1)}_2 ={}& \int\! d^{D\!}x \, \phi^2 (\partial_\mu\phi) (\partial^\mu\phi) \, ,
\label{Ss(1)2}
\\
S^{(1)}_3 ={}& \int\! d^{D\!}x \, (\square\phi) (\square\phi) \, ,
\label{Ss(1)3}
\end{align}
the second order correction,
\begin{equation}
\epsilon^2 S^{(2)} 
	= \epsilon^2 \sum_{n=1}^{6} \alpha^{(2)}_n S^{(2)}_n \, ,
\end{equation}
out of all possible independent dimension 8 terms,
\begin{align}
S^{(2)}_1 ={}& \int\! d^{D\!}x \, \phi^8 \, ,
\label{Ss(2)1}
\\
S^{(2)}_2 ={}& \int\! d^{D\!}x \, \phi^4 (\partial_\mu\phi) (\partial^\mu\phi) \, ,
\label{Ss(2)2}
\\
S^{(2)}_3 ={}& \int\! d^{D\!}x \, (\partial_\mu\phi) (\partial^\mu\phi) 
	(\partial_\nu\phi) (\partial^\nu\phi) \, ,
\label{Ss(2)3}
\\
S^{(2)}_4 ={}& \int\! d^{D\!}x \, 
	\phi (\partial_\mu\phi) (\partial^\mu\phi) (\square\phi) \, ,
\label{Ss(2)4}
\\
S^{(2)}_5 ={}& \int\! d^{D\!}x \, \phi^2 (\square\phi) (\square\phi) \, ,
\label{Ss(2)5}
\\
S^{(2)}_6 ={}& \int\! d^{D\!}x \, \phi^2 (\partial_\mu \square \phi) 
	(\partial^\mu \square \phi) \, ,
\label{Ss(2)6}
\end{align}
and the third order correction,
\begin{equation}
\epsilon^3 S^{(3)} 
	= \epsilon^3 \sum_{n=1}^{13} \alpha^{(3)}_n S^{(3)}_n \, ,
\end{equation}
out of all possible independent dimension 10 terms,
\begin{align}
S^{(3)}_1 ={}& \int\! d^{D\!}x \, \phi^{10} \, ,
\label{Ss(3)1}
\\
S^{(3)}_2 ={}& \int\! d^{D\!}x \, \phi^{6} (\partial_\mu \phi) (\partial^\mu \phi) \, ,
\label{Ss(3)2}
\\
S^{(3)}_3 ={}& \int\! d^{D\!}x \, \phi^{2} (\partial_\mu \phi) (\partial^\mu \phi)
	(\partial_\nu \phi) (\partial^\nu \phi) \, ,
\label{Ss(3)3}
\\
S^{(3)}_4 ={}& \int\! d^{D\!}x \, \phi^3 (\partial_\mu \phi) (\partial^\mu \phi) 
	(\square\phi) \, ,
\label{Ss(3)4}
\\
S^{(3)}_5 ={}& \int\! d^{D\!}x \, \phi^4 (\square \phi) (\square \phi) \, ,
\label{Ss(3)5}
\\
S^{(3)}_6 ={}& \int\! d^{D\!}x \, (\partial_\mu \phi) (\partial^\mu \phi)
	(\square \phi) (\square \phi) \, ,
\label{Ss(3)6}
\\
S^{(3)}_{7} ={}& \int\! d^{D\!}x \, (\partial_\mu \phi) (\partial^\mu \phi)
	(\partial_\nu \partial_\rho \phi) (\partial^\nu \partial^\rho \phi) \, ,
\label{Ss(3)7}
\\
S^{(3)}_{8} ={}& \int\! d^{D\!}x \, (\partial_\mu\phi) (\partial_\nu \phi)
	(\partial^\mu \partial^\nu \phi) (\square\phi) \, ,
\label{Ss(3)8}
\\
S^{(3)}_{9} ={}& \int\! d^{D\!}x \, \phi (\square \phi) (\square \phi) (\square\phi) \, ,
\label{Ss(3)9}
\\
S^{(3)}_{10} ={}& \int\! d^{D\!}x \, \phi (\partial_\mu\partial_\nu \phi)
	(\partial^\mu\partial^\nu \phi) (\square\phi) \, ,
\label{Ss(3)10}
\\
S^{(3)}_{11} ={}& \int\! d^{D\!}x \, \phi (\partial_\mu\partial_\nu \phi)
	(\partial^\nu\partial^\rho \phi) (\partial_\rho \partial^\mu \phi) \, ,
\label{Ss(3)11}
\\
S^{(3)}_{12} ={}& \int\! d^{D\!}x \, \phi^2 (\partial_\mu \square \phi)
	(\partial^\mu \square \phi) \, ,
\label{Ss(3)12}
\\
S^{(3)}_{13} ={}& \int\! d^{D\!}x \, (\square \square \phi) (\square \square \phi) \, .
\label{Ss(3)13}
\end{align}
The EFT expansion introduces one higher derivative term~(\ref{Ss(1)3})
at linear order, three higher derivative terms~(\ref{Ss(2)4}-\ref{Ss(2)6}) at
second order, and ten~(\ref{Ss(3)4}-\ref{Ss(3)13}) at third order. 
This is a rather generic feature of the EFT expansion, and it is hard to come
up with a symmetry principle that would exclude them. These higher derivative 
terms do not mean that at each order in $\epsilon$ new degrees of freewhatdom
appear. It is rather the assumed approximation scheme that represents corrections
as local terms that is to blame for the seeming introduction of extra degrees 
of freedom. These degrees are spurious, and ought to be treated in the
same spirit in which they arise -- perturbatively, which actually eliminates them.

The following subsections analyze the reduction of derivative order 
at successive orders in the expansion parameter $\epsilon$ using three
methods -- reduction at the equation of motion level, reduction via
perturbative constraints method of section~\ref{sec: Formalism},
and reduction via field redefinitions. Considering 
all the possible corrections~(\ref{Ss(1)1}-\ref{Ss(3)13}) in 
is not necessary for the points that we want to make, and would only 
obscure them with the sheer number of formulas without introducing anything
essentially new to the discussion. Rather we will only consider
the highest derivative term at each order~(\ref{Ss(1)3}),~(\ref{Ss(2)6}), 
and~(\ref{Ss(3)13}), and in addition~(\ref{Ss(3)7}).
 The emphasis in this section is not on the exact
form of the reduced theories, but rather on the lessons that can be
learned about methods from considering simpler examples.

It is worth noting that in~\cite{Crisostomi:2017aim} the Lorentz invariant
systems of multiple interacting scalar fields with actions containing 
irreducible second derivatives were analysed, and classified according whether
they can be brought to explicit lower order form, and whether this 
preserves manifest Lorentz covariance.


\subsection{Derivative reduction at first order}

Let us consider the the action~(\ref{leading order action}) with the first order
correction~(\ref{Ss(1)3}),
\begin{equation}
S = \int\! d^{D\!}x \, \biggl\{ - \frac{1}{2} (\partial_\mu\phi) (\partial^\mu\phi)
	- U + \epsilon \alpha_1 (\square\phi) (\square\phi) \biggr\} \, ,
\label{first corrected action}
\end{equation}
while neglecting all the contributions of order $\mathcal{O}(\epsilon^2)$. 
We will be looking to reduce the derivative order of this system,
and represent it with an equivalent lower derivative system that captures
correctly the solutions to order $\epsilon$. Three methods are considered
in the following: reducing the derivative order at the level of the equation of motion,
reducing the derivative order by implementing perturbative second class
constraints of Section~\ref{sec: Formalism}, and reducing the order via
field redefinition.


\subsubsection{Reduction at the equation of motion level}

The equation of motion descending from the action~(\ref{first corrected action}) is
\begin{equation}
\square\phi - U' + 2\epsilon\alpha_1 \square\square\phi = 0 \, .
\label{first order unreduced equation}
\end{equation}
Applying the leading order equation of motion to the first order term is 
straightforward and yields
\begin{equation}
\square \phi - U' + 2\epsilon\alpha_1 \bigl[ U'' \square\phi 
	+ U''' (\partial_\mu\phi) (\partial^\mu\phi) \bigr] = 0 \, ,
\end{equation}
which correctly reproduces the perturbative solutions of the original 
equation~(\ref{first order unreduced equation}) to linear order in $\epsilon$.
But this is not the only lower (time) derivative equation that does it. In fact,
since we can always use $\square\phi = U'$ at linear order, there is a one-parameter 
family of equations that accomplish this task,
\begin{equation}
\square\phi - U' + 2\epsilon \alpha_1 \bigl[ (1\!+\!b) U'' \square\phi
	+ U'''(\partial_\mu\phi) (\partial^\mu\phi) - b \, U'' U' \bigr] = 0 \, ,
\label{first order b equation}
\end{equation}
where $b$ is the free parameter. There is one distinguished choice for this parameter 
$b\!=\!1$, for which there is an action formulation of the equation of motion,
in the sense that the equation following from the action is exactly the one for
$b\!=\!1$ with no further manipulations required. We will derive this action in 
the following subsection via the method of perturbative second-class constraints.
As noted in the Introduction, an action formulation allows for resummation
that conserves exactly  the charges associated to global symmetries.


\subsubsection{Reduction via perturbative constraints}

Here we apply the method described in Sec.~\ref{sec: Formalism} to 
reduce the derivative order of the action~(\ref{first corrected action}).
The starting point is the definition of the extended action, which 
is obtained from the configuration space one
by promoting time derivatives to independent fields,
\begin{equation}
\dot{\phi}\rightarrow V_1, 
\qquad
- \square\phi = \ddot{\phi} - \nabla^2\phi = \dot{V}_1 - \nabla^2\phi \rightarrow V_2 \, ,
\end{equation}
and
by introducing appropriate Lagrange multipliers $\pi$ and $\pi_1$ to ensure 
the on-shell equivalence,
\begin{align}
\mathcal{S} 
={}& \int\! d^{D\!}x \, \biggl\{ \frac{1}{2} V_1^2 - \frac{1}{2}(\nabla\phi)^2 - U
	+ \epsilon \alpha_1 V_2^2
	+ \pi (\dot{\phi} - V_1) + \pi_1 (\dot{V}_1 - \nabla^2\phi -V_2) \biggr\}
\nonumber \\
={}& \int\! d^{D\!}x \, \biggl\{
	\pi \dot{\phi} + \pi_1 \dot{V}_1 - \biggl[
		\pi V_1 + \pi_1 ( V_2 + \nabla^2\phi) 
		- \frac{1}{2} V_1^2 + \frac{1}{2}(\nabla\phi)^2 + U
		- \epsilon \alpha_1 V_2^2
		\biggr]
	\biggr\} \, .
\label{first order extended action}
\end{align}
Proceeding with the recipe given in Sec.~\ref{sec: Formalism},
the equations resulting from the action principle with respect to all the independent
fields are
\begin{eqnarray}
&\displaystyle \frac{\delta \mathcal{S}}{\delta V_2} \approx 0
\qquad \Rightarrow \qquad&
\pi_1 \approx 2 \epsilon \alpha_1 V_2
\\
&\displaystyle \frac{\delta \mathcal{S}}{\delta V_1} \approx 0
\qquad \Rightarrow \qquad&
\dot{\pi}_1 \approx - \pi + V_1 \, ,
\\
&\displaystyle \frac{\delta \mathcal{S}}{\delta \phi} \approx 0
\qquad \Rightarrow \qquad&
\dot{\pi} \approx\nabla^2\phi - U' - \nabla^2\pi_1 \, ,
\\
&\displaystyle \frac{\delta \mathcal{S}}{\delta \pi} \approx 0
\qquad \Rightarrow \qquad&
\dot{\phi} \approx V_1 \, ,
\\
&\displaystyle \frac{\delta \mathcal{S}}{\delta \pi_1} \approx 0
\qquad \Rightarrow \qquad&
\dot{V}_1 \approx V_2 + \nabla^2\phi \, .
\end{eqnarray}
Multiplying these equations by $\epsilon$ and discarding the 
terms~$\mathcal{O}(\epsilon^2)$ gives us the leading order equations,
\begin{align}
\epsilon \pi_1 \approx{}& 0 \, ,
\label{first order constraint}
\\
\epsilon \dot{\pi}_1 \approx{}& \epsilon (- \pi + V_1 ) \,  ,
\\
\epsilon \dot{\pi} \approx{}& \epsilon (\nabla^2\phi - U' - \nabla^2\pi_1 ) \, ,
\\
\epsilon \dot{\phi} \approx{}& \epsilon V_1 \, ,
\\
\epsilon \dot{V}_1 \approx{}& \epsilon (V_2 + \nabla^2\phi ) \, ,
\label{first order dotV1}
\end{align}
where now it is clear that Eq.~(\ref{first order constraint}) takes the
form of a primary constraint, and $V_2$ is its associated Lagrange multiplier. 
Requiring the consistency of the primary constraint
(vanishing of its time derivative) generates a secondary constraint
\begin{equation}
\epsilon \dot{\pi}_1 \approx \epsilon (- \pi + V_1) \approx 0 \, 
\qquad \Rightarrow \qquad \epsilon V_1 \approx \epsilon \pi \, .
\end{equation}
The two constraint are second-class, and determine the Lagrange multiplier of 
the primary constraint,
\begin{equation}
\epsilon \frac{d}{dt} (-\pi + V_1) \approx \epsilon (U' + V_2) \approx 0 
\qquad \Rightarrow \qquad
\epsilon V_2 \approx - \epsilon U' \, .
\label{leading order V2}
\end{equation}
This determines $\pi_1, V_1, V_2$ to leading order depending algebraically
of $\phi$ and $\pi$.

After solving for the leading order constraints, we turn to the first order 
equations~(\ref{first order constraint}-\ref{first order dotV1}). The first of these
equations again takes the form of the primary constraint, because upon using
the leading order solution of the constraint equations~(\ref{leading order V2}) it reads
\begin{equation}
\pi_1 \approx - 2\epsilon \alpha_1 U' \, .
\end{equation}
In the same way as for the leading order, this primary constraint generates 
a secondary constraint, the two of them being second-class, and determining the
Lagrange multiplier,
\begin{align}
V_1 \approx{}& \pi - 2\epsilon \alpha_1 U'' \pi \, ,
\\
V_2 \approx{}& - U' - 2 \epsilon\alpha_1 \Bigl\{ U''' \bigl[\pi^2 - (\nabla\phi)^2 \bigr]
	- U'' U' \Bigr\} \, .
\end{align}
We have solved for $\pi_1$, $V_1$, and $V_2$ to first order as functions of
$\pi$ and $\phi$. Since these relations are algebraic, in the sense that they do not
involve any time derivatives, we can promote them to strong (off-shell) equalities,
and plug them back into the extended action~(\ref{first order extended action})
to obtain the phase space reduced action, which after a couple of partial integrations
reads
\begin{equation}
\mathscr{S}_{\rm red} = \int\! d^{D\!}x \, \biggl\{
	\bigl( 1 + 2\epsilon \alpha_1 U'' \bigr) \pi \dot{\phi} 
	- \biggl[ \frac{1}{2} \pi^2 + \frac{1}{2} (\nabla\phi)^2 + U
	+ \epsilon \alpha_1 \bigl[ 2 U'' (\nabla\phi)^2 + (U')^2 \bigr]
		 \biggr]
	\biggr\} \, .
\label{phase action reduced}
\end{equation}
In fact, because of the structure of the extended action, we do not actually
need to solve for velocities in this example beyond leading order since
their higher orders cancel when plugged into the action.

Note that the  phase space action~(\ref{phase action reduced}) obtained
via the imposition of perturbative second-class constraints is not in a 
canonical form, due to the symplectic
part not being just $\pi \dot{\phi}$, but is modified by the field dependent prefactor.
This modified symplectic part means that the the Poissons bracket between $\phi$
and $\pi$ is not canonical, but rather
\begin{equation}
 \bigl\{ \phi (t,\vec{x}), \pi(t,\vec{x}^{\,\prime}) \bigr\} 
	= \frac{\delta^{D-1}(\vec{x}\!-\!\vec{x}^{\,\prime})}{1 + 2\epsilon \alpha_1 U''}
	= \bigl[ 1 - 2\epsilon \alpha_1 U'' \bigr] \delta^{D-1}(\vec{x}\!-\!\vec{x}^{\,\prime}) \, .
\end{equation}
This is not surprising, since such a bracket results from solving the second-class
constraints, and is in fact nothing else then the Dirac bracket.

Even though it is not necessary, the symplectic part of the phase space action 
can be canonicalized by a non-canonical transformation of the conjugate momentum 
field,
\begin{equation}
\pi \rightarrow \bigl( 1 - 2\epsilon \alpha_1 U'' \bigr) \pi \, ,
\label{momentum transformation}
\end{equation}
after which the phase space action takes the form
\begin{equation}
\mathscr{S}_{\rm red} = \int\! d^{D\!}x \, \biggl\{
	\pi \dot{\phi}
	- \biggl[ \frac{1}{2} \pi^2 + \frac{1}{2} (\nabla\phi)^2 + U
	+ \epsilon \alpha_1 \Bigl( 2 U'' [- \pi^2 + (\nabla\phi)^2 ] + (U')^2 \Bigr) \biggr] 
	\biggr\} \, .
\label{first order canonical}
\end{equation}
The transformation of the conjugate momentum 
field~(\ref{momentum transformation}) is not the unique transformation
that puts the phase space action into a canonical form. This can also be accomplished by 
transforming the scalar field itself,
\begin{equation}
\phi \rightarrow \phi - 2\epsilon \alpha_1 U' \, .
\end{equation}
We are hesitant to make field transformations in the phase space action.
Even though it is not the case in this specific example, it is very easy to lose
manifest Lorentz covariance when making a field transformation that involves
gradients of the field, or the conjugate momentum.

The reduced configuration space action is obtained from the phase space 
one~(\ref{first order canonical}) in the usual manner. First we solve 
the Hamilton's equation
for the conjugate momentum filed in terms of the scalar and its time derivative,
\begin{equation}
\frac{\delta \mathscr{S}_{\rm red}}{\delta \pi} \approx 0
\qquad \Rightarrow \qquad
\pi \approx \dot{\phi} - 4 \epsilon \alpha_1 U'' \dot{\phi} \, .
\end{equation}
This relation is then promoted to the on-shell equality and plugged into the
phase space action to obtain the configuration space one, which can be written in a 
manifestly Lorentz invariant form
\begin{equation}
S_{\rm red} = \int\! d^{D\!}x \, \biggl\{
	- \frac{1}{2} \bigl[ 1 + 4\epsilon \alpha_1 U'' \bigr] 
		(\partial_\mu\phi) (\partial^\mu\phi) - U - \epsilon \alpha_1 (U')^2
	\biggr\} \, .
\label{first order configuration action}
\end{equation}
The effect of the higher derivative correction when reduced is to
renormalize the kinetic term (by a field-dependent factor), and to
add another potential term.
The equation of motion following from the reduced lower derivative action is
\begin{equation}
\square\phi - U' + 2\epsilon \alpha_1 \bigl[ 2 U'' (\square\phi)
	+ U''' (\partial_\mu\phi) (\partial^\mu\phi) - U'' U' \bigr] \, ,
\end{equation}
which is precisely the $b\!=\!1$ equation~(\ref{first order b equation}) 
we have obtained by reducing the
derivative order at the level of the equation of motion.

We have derived the lower derivative action~(\ref{first order configuration action})
that captures completely to first order in $\epsilon$ the subspace of 
perturbatively expandable solutions of the
higher derivative action~(\ref{first corrected action}).


\subsubsection{Derivative reduction via field redefinition}

Since the higher derivative terms in the leading
order action~(\ref{first corrected action})
are assumed to arise perturbatively, and are to be treated perturbatively, we
may employ field redefinitions that include higher derivatives. It is
simple to see that the following redefinition
\begin{equation}
\phi \rightarrow \phi - \epsilon \alpha_1 \square \phi \, ,
\end{equation}
removes all the higher derivatives to order $\mathcal{O}(\epsilon^2)$,
\begin{equation}
S \rightarrow \int\! d^{D\!}x \, \biggl\{
	- \frac{1}{2} \bigl[ 1 + 2\alpha_1\epsilon U'' \bigr] 
		(\partial_\mu\phi) (\partial^\mu \phi) - U
	\biggr\} \, ,
\end{equation}
accomplishing the desired task. This is not the same action obtained by
the method of perturbative constraints~(\ref{first order configuration action}).
The discrepancy is resolved by noting that here we are still left with the freedom
of field redefinitions not including derivatives, which do not change the derivative
order of the action. In particular, an additional redefinition,
\begin{equation}
\phi \rightarrow \phi + \epsilon \alpha_1 U' \, ,
\end{equation}
transforms the action exactly into~(\ref{first order configuration action}),
\begin{equation}
S \rightarrow \int\! d^{D\!}x \, \biggl\{
	- \frac{1}{2} \bigl[ 1 + 4\alpha_1\epsilon U'' \bigr] 
		(\partial_\mu\phi) (\partial^\mu \phi) - U - \epsilon\alpha_1 (U')^2
	\biggr\} \, .
\end{equation}
This demonstrates the equivalence of field redefinitions to the other 
two derivative reduction methods.


\subsection{Derivative reduction at second order}

At the second order in $\epsilon$ we consider only the two higher derivative
corrections to the leading order 
action~(\ref{leading order action}), the first one~(\ref{Ss(1)3}) 
arising at linear level,
and the second one~(\ref{Ss(2)6})  at the quadratic,
\begin{equation}
S = \int\! d^{D\!}x \, \biggl\{
	- \frac{1}{2} (\partial_\mu \phi) (\partial^\mu \phi) - U
	+ \epsilon \alpha_1 (\square\phi) (\square\phi)
	+ \epsilon^2 \alpha_2 (\partial_\mu \square\phi) (\partial^\mu\square\phi)
	\biggr\} \, .
\label{2nd action}
\end{equation}
%


\subsubsection{Reduction at the equation of motion level}

The equation of motion following from the starting higher derivative action is
\begin{equation}
\square \phi - U' + 2 \epsilon \alpha_1 \square\square\phi
	- 2 \epsilon^2 \alpha_2 \square\square\square \phi = 0 \, .
\end{equation}
By successive applications of the lower order equations in the higher
order terms this equation can be reduced to the second order one in derivatives,
\begin{align}
\MoveEqLeft[3]
\square\phi - U'
	+ 2\epsilon\alpha_1 \bigl[ U'' U' + U'''(\partial_\mu\phi) (\partial^\mu\phi) \bigr]
	- 2\epsilon^2(\alpha_2 + 2\alpha_1^2) \Bigl\{
	[U''U']' U' 
\nonumber \\
&	+\bigl[ 2 U^{(4)} U' + 5 U''' U'' \bigr] (\partial_\mu\phi) (\partial^\mu\phi)
	+ U^{(5)} (\partial_\mu\phi) (\partial^\mu\phi) (\partial_\nu\phi) (\partial^\nu\phi)
\nonumber \\
&	+ 4 U^{(4)} (\partial^\mu\phi) (\partial^\nu\phi) (\partial_\mu\partial_\nu\phi)
	+ 2 U''' (\partial_\mu\partial_\nu\phi) (\partial^\mu\partial^\nu \phi)
	\Bigr\} = 0 \, .
\end{align}
There is no obstacle to accomplishing this task.
Just as in the first order case, this is not a unique equation capturing solutions
to the second order in $\epsilon$, and only particular forms of the reduced
equation have action formulations.


\subsubsection{Reduction via perturbative constraints}

The starting point of the method of derivative reduction via
perturbative constraints is the definition of the extended action.
It is obtained by promoting
time derivatives to independent fields,
\begin{equation}
\dot{\phi}\rightarrow V_1, 
\qquad
-\square\phi =  \dot{V}_1 - \nabla^2\phi \rightarrow V_2\, ,
\qquad
-\partial_0 \square\phi = \dot{V}_2 \rightarrow V_3,
\end{equation}
and introducing Lagrange multipliers $\pi, \pi_1, \pi_2$ to ensure on-shell equivalence,

\begin{align}
\mathcal{S} ={}& \int\! d^{D\!}x \, \biggl\{
	\frac{1}{2} V_1^2 - \frac{1}{2} (\nabla\phi)^2 - U
	+ \epsilon\alpha_1 V_2^2
	- \epsilon^2 \alpha_2 V_3^2 + \epsilon^2 \alpha_2 (\nabla^2V_1)^2
\nonumber \\
&	\hspace{4cm}
	+ \pi(\dot{\phi} - V_1) + \pi_1 (\dot{V}_1 - \nabla^2\phi - V_2)
	+ \pi_2 (\dot{V}_2 - V_3)
	\biggr\}
\nonumber \\
={}& \int\! d^{D\!}x \, \biggl\{ 
	\pi\dot{\phi} + \pi_1 \dot{V}_1 + \pi_2 \dot{V}_2
	- \biggl[
	\pi V_1 + \pi_1(\nabla^2\phi+V_2) + \pi_2 V_3
\nonumber \\
&	\hspace{3cm}
	- \frac{1}{2}V_1^2 + \frac{1}{2} (\nabla\phi)^2 + U
	- \epsilon\alpha_1 V_2^2
	+ \epsilon^2 \alpha_2 [V_3^2 - (\nabla V_2)^2]
	\biggr]
	\biggr\} \, .
\label{2nd extended}
\end{align}
The procedure here mirrors the one for the first order case.
We provide all the steps in order to illustrate the method 
better. The equations resulting from the action principle 
with respect to all the independent fields are
\begin{eqnarray}
&\displaystyle \frac{\delta\mathcal{S}}{\delta V_3} \approx 0
\qquad \Rightarrow \qquad &
\pi_2 \approx - 2\epsilon^2\alpha_2 V_3 \, ,
\label{fe}
\\
&\displaystyle \frac{\delta\mathcal{S}}{\delta V_2} \approx 0
\qquad \Rightarrow \qquad &
\dot{\pi}_2 \approx - \pi_1 + 2\epsilon\alpha_1 V_2
	- 2\epsilon^2 \alpha_2 \nabla^2 V_2 \, ,
\\
&\displaystyle \frac{\delta\mathcal{S}}{\delta V_1} \approx 0
\qquad \Rightarrow \qquad &
\dot{\pi}_1 \approx - \pi  V_1 \, ,
\\
&\displaystyle \frac{\delta\mathcal{S}}{\delta \phi} \approx 0
\qquad \Rightarrow \qquad &
\dot{\pi} \approx - \nabla^2 \pi_1 + \nabla^2\phi - U' \, ,
\\
&\displaystyle \frac{\delta\mathcal{S}}{\delta \pi} \approx 0
\qquad \Rightarrow \qquad &
\dot{\phi} \approx V_1 \, ,
\\
&\displaystyle \frac{\delta\mathcal{S}}{\delta \pi_1} \approx 0
\qquad \Rightarrow \qquad &
\dot{V}_1 \approx \nabla^2\phi + V_2 \, ,
\\
&\displaystyle \frac{\delta\mathcal{S}}{\delta \pi_2} \approx 0
\qquad \Rightarrow \qquad &
\dot{V}_2 \approx V_3 \, .
\label{le}
\end{eqnarray}
Multiplying these equations by $\epsilon^2$ gives the leading order 
equations,
\begin{align}
\epsilon^2 \pi_2 \approx{}& 0 \, ,
\\
\epsilon^2 \dot{\pi}_2 \approx{}& - \epsilon^2 \pi_1 \, ,
\\
\epsilon^2 \dot{\pi}_1 \approx{}& \epsilon^2 (-\pi + V_1) \, ,
\\
\epsilon^2 \dot{\pi} \approx{}&  \epsilon^2 (-\nabla^2\pi_1 + \nabla^2\phi - U') \, ,
\\
\epsilon^2 \dot{\phi} \approx{}& \epsilon^2 V_1 \, ,
\\
\epsilon^2 \dot{V}_1 \approx{}&	\epsilon^2 (\nabla^2\phi+V_2) \, ,
\\
\epsilon^2 \dot{V}_2 \approx{}& \epsilon^2 V_3 \, .
\end{align}
The first of these is a constraint equation. Requiring its time 
derivative to vanish on-shell produces another constraint,
\begin{equation}
\epsilon^2 \dot{\pi}_2 \approx - \epsilon^2 \pi_1 \approx 0 \, .
\end{equation}
Consistency of this constraint produces the next one,
\begin{equation}
\epsilon^2 \dot{\pi}_2 \approx \epsilon^2 (- \pi + V_1) \approx 0
\qquad \Rightarrow \qquad 
\epsilon^2 V_1 \approx \epsilon^2 \pi \, ,
\end{equation}
Again requiring consistency produces a constraint,
\begin{equation}
\epsilon^2 \frac{d}{dt} (- \pi + V_1) \approx
	\epsilon^2 (V_2 + U') \approx 0
\qquad \Rightarrow \qquad 
\epsilon^2 V_2 \approx - \epsilon^2 U' \, .
\end{equation}
Finally, the consistency of this constraint determines the Lagrange multiplier,
\begin{equation}
\epsilon^2 \frac{d}{dt} (V_2 + U') \approx \epsilon^2 (V_3 + U'' \pi) \approx 0
\qquad \Rightarrow \qquad 
\epsilon^2 V_3 \approx - \epsilon^2 U'' \pi \, .
\end{equation}

After solving the leading order constraints we proceed to the fist order ones.
Multiplying equations~(\ref{fe}-\ref{le}) by $\epsilon$, and making use of the 
leading order constraints gives the equations at first order
\begin{align}
\epsilon \pi_2 \approx{}& 0 \, ,
\\
\epsilon \dot{\pi}_2 \approx{}& - \epsilon\pi_1 - 2\epsilon^2\alpha_1U' \, ,
\\
\epsilon \dot{\pi}_1 \approx{}& \epsilon (-\pi + V_1) \, ,
\\
\epsilon \dot{\pi} \approx{}& \epsilon (-\nabla^2\pi_1 + \nabla^2\phi - U') \, ,
\\
\epsilon \dot{\phi} \approx{}& \epsilon V_1 \, ,
\\
\epsilon \dot{V}_1 \approx{}& \epsilon (\nabla^2\phi+V_2) \, ,
\\
\epsilon \dot{V}_2 \approx{}& \epsilon V_3 \, .
\end{align}
The first of these equations is a constraint. Its consistency generates
another constraint,
\begin{equation}
\epsilon \dot{\pi}_2 \approx - \epsilon \pi_1  -2\epsilon^2 \alpha_1 U' \approx 0
\qquad \Rightarrow \qquad 
\epsilon \pi_1 \approx - 2\epsilon^2 U' \, .
\end{equation}
Its consistency generates the next constraint,
\begin{equation}
\epsilon \frac{d}{dt} (\pi_1 + 2\epsilon\alpha_1 U') \approx 0
\qquad \Rightarrow \qquad 
\epsilon V_1 \approx \epsilon \pi - 2\epsilon^2\alpha_1 U'' \pi \, ,
\end{equation}
which by consistency requirements generates yet another one
\begin{equation}
\epsilon \frac{d}{dt} (V_1 - \pi + 2\epsilon\alpha_1 U'' \pi) \approx 0
\qquad \Rightarrow \qquad 
\epsilon V_2 \approx - \epsilon U'  
	+2\epsilon^2\alpha_1 \bigl( U'''[-\pi^2+(\nabla\phi)^2] + U''U' \bigr) \, .
\end{equation}
Finally consistency of this constraint determines $V_3$ at first order,
but we do not require this to construct the reduced action.

In the last stage of solving the constraints we turn to the second
order equations~(\ref{fe}-\ref{le}). By successively requiring consistency
of constraints solve the perturbative constraints fully. Here we give
just the terms that do not cancel when plugged into the action,
\begin{align}
\pi_2 \approx{}& 2 \epsilon^2 \alpha_2 U'' \pi \, ,
\\
\pi_1 \approx{}& - 2\epsilon \alpha_1 U' 
	+ 2\epsilon^2(\alpha_2+2\alpha_1^2)
		\bigl( U'''[-\pi^2+(\nabla\phi)^2] + U''U' \bigr) \, ,
\\
V_1 \approx{}& \pi - 2\epsilon\alpha_1 U'' \pi + \mathcal{O}(\epsilon^2) \, ,
\\
V_2 \approx{}& - U' + \mathcal{O}(\epsilon) \, ,
\\
V_3 \approx{}& - U'' \pi + \mathcal{O}(\epsilon) \, .
\end{align}
These are now plugged into the extended action~(\ref{2nd extended}) as
strong equalities, resulting in the reduced phase space action, which after some
partial integrations can be written as
\begin{align}
\mathscr{S}_{\rm red} ={}& \int\! d^{D\!}x \, \biggl\{
	\biggl[ \pi + 2\epsilon\alpha_1 U'' \pi
	+ 2\epsilon^2 (\alpha_2 + 2\alpha_1^2) \Bigl( \frac{1}{3} U^{(4)} \pi^3
	- U^{(4)} (\nabla\phi)^2 \pi 
\nonumber \\
&	\hspace{6cm}
	+ 2 \nabla(U'''\pi\nabla\phi) - [U'''U' + 2(U'')^2] \pi \Bigr) \biggr] \dot{\phi}
\nonumber \\
&	- \biggl[ \frac{1}{2} \pi^2 + \frac{1}{2}(\nabla\phi)^2 + U
	+ \epsilon \alpha_1 \bigl( 2 U''(\nabla\phi)^2 + (U')^2 \bigr) 
	- \alpha_2 \epsilon^2 (U'')^2 (\nabla\phi)^2
\nonumber \\
&	+ 2\epsilon^2 (\alpha_2 + 2\alpha_1^2) 
	\bigl( U'''[-\pi^2 + (\nabla\phi)^2] + U''U' \bigr) (\nabla^2\phi - U')
	-(\alpha_2 + 2\alpha_1^2) \epsilon^2 (U'')^2 \pi^2 \biggr]
	\biggr\} \, .
\label{2nd phase action}
\end{align}
The reduced phase space action at second order comes with the noncanonical
symplectic part, but such that it can be canonicalized by redefining the
conjugate momentum field, just as we had at the first order reduction.
We do not bother with canonicalizing the phase space action, but proceed
to construct the reduced configuration space action. First we solve for the
conjugate momentum field
\begin{equation}
\frac{\delta\mathscr{S}_{\rm red}}{\delta \pi} \approx 0
\qquad \Rightarrow \qquad
\pi \approx \dot{\phi} + 2\epsilon \alpha_1 U'' \dot{\phi} + \mathcal{O}(\epsilon^2) \, ,
\end{equation}
and then plug it into the phase space action~(\ref{2nd phase action}). 
Thus we obtain the reduced configuration space action, which after some number
of partial integrations and groupings of the terms can be written in a
manifestly Lorentz invariant form,
\begin{align}
S_{\rm red} ={}& \int\! d^{D\!}x \, \biggl\{
	- \frac{1}{2} \Bigl[ 1 + 4\epsilon \alpha_1 U''
		- 8(\alpha_2 + 2\alpha_1^2) U''' U'
		- 2(3\alpha_2 + 4\alpha_1^2) (U'')^2 \Bigr] 
		(\partial_\mu\phi) (\partial^\mu\phi)
\nonumber \\
&	\hspace{0.3cm}
	- U - \epsilon \alpha_1 (U')^2
	+ 2(\alpha_2 + 2\alpha_1^2) U'' (U')^2
	- 2(\alpha_2 + 2\alpha_1^2) U''' (\partial_\mu\phi) (\partial^\mu\phi) (\square \phi)
	\biggr\} \, .
\label{II reduced}
\end{align}
Apart from renormalizing the kinetic term, and modifying the potential 
already encountered at first order, the effect of reduction the considered 
higher derivative  is also to introduce a new term -- the last
one in the action above. This term technically is a higher derivative one 
 due to the presence of d'Alembertian operator, but it is not a genuine 
 higher derivative terms, since the higher time derivative in it 
 can be removed by partial integration,
and are only introduced for the sake of manifest Lorentz invariance. 
Moreover, the very method used to derive it guarantees no higher derivatives 
in the final result. It is
worth noting that derivative reduction introduces new types of terms 
containing four derivatives into the reduced action, but the kind
that are not genuine higher derivative ones.


\subsubsection{Derivative reduction via field redefinition}

With a bit of work one can see that the field transformation,
\begin{equation}
\phi \rightarrow \phi - \epsilon \alpha_1 \square\phi
	+ \frac{3}{2} \epsilon^2 \alpha_1^2 \square \square \phi
	+ 2\epsilon^2 \alpha_1^2 U'' \square\phi
	+ 2 \epsilon^2 \alpha_2 \square \bigl[ \square\phi + U' \bigr] \, ,
\end{equation}
removes all the genuine higher derivative terms from the configuration space 
action~(\ref{2nd action}), and transforms it into
\begin{align}
S \rightarrow{}& \int\! d^{D\!}x \, \biggl\{
	- \frac{1}{2} \Bigl[ 1 + 2\epsilon\alpha_1 U''
		- 2\epsilon^2(\alpha_2 + 2\alpha_1^2)(U'')^
		- 4\epsilon^2 \alpha_1^2 U''' U' \Bigr] 
		(\partial_\mu\phi) (\partial^\mu\phi) 
\nonumber \\
&	\hspace{2cm}
	- U - \frac{3}{2} \epsilon^2 \alpha_1^2 U''' U' 
		(\partial_\mu\phi) (\partial^\mu\phi)  (\square\phi)
	\biggr\} \, .
\end{align}
This action is not in the form of the reduced one~(\ref{II reduced}) obtained
via perturbative constraints method. But again, this is not an inconsistency, since
an additional field redefinition
\begin{equation}
\phi \rightarrow \phi 
	+ \epsilon \alpha_1 U' 
	- \frac{5}{2} \epsilon^2 \alpha_1^2 U''' (\partial_\mu\phi) (\partial^\mu\phi)
	- \frac{9}{2} \epsilon^2 \alpha_1^2 U'' U'
	- 2\epsilon^2\alpha_2 U''' (\partial_\mu\phi) (\partial^\mu\phi)
	- 2\epsilon^2 \alpha_2 U''U' \, ,
\end{equation}
bring the above action precisely to the form of~(\ref{II reduced}). The field
redefinition method is consistent with the perturbative constraints method.


\subsection{Derivative reduction at third order -- Part I}

In order to illustrate the points made at this section we consider the leading
order action~(\ref{leading order action}) with the only nonvanishing higher
order correction~(\ref{Ss(3)13}) coming in at the third order EFT expansion,
\begin{equation}
S = \int\! d^{D\!}x \, \biggl\{
	- \frac{1}{2} (\partial_\mu\phi) (\partial^\mu\phi) - U
	+ \epsilon^3 \alpha_3 (\square\square\phi) (\square\square\phi) 
	\biggr\} \, .
\label{third order HD action}
\end{equation}
At this order we make the point that field redefinitions are necessary in order to 
define an equivalent lower derivative, manifestly Lorentz invariant, 
configuration space action formulation of the 
higher derivative system.


\subsubsection{Reduction at the equation of motion level}

The equation of motion descending from the action~(\ref{third order HD action}) is
\begin{equation}
\square \phi - U' + 2 \epsilon^3 \alpha_3 \square\square\square\square\phi = 0 \, .
\end{equation}
Just as in the cases of the first and the second order reduction, we make use
of the lower order equation to reduce the derivative order of the higher order
part of the equation. The resulting equation takes the form
\begin{align}
\MoveEqLeft[4]
\square\phi - U' + 2\epsilon^3 \alpha_3 \biggl\{
	A(\phi) + B(\phi) \times (\partial_\mu\phi) (\partial^\mu\phi)
	+ C(\phi) \times (\partial_\mu\phi) (\partial^\mu\phi) 
			(\partial_\nu\phi) (\partial^\nu\phi)
\nonumber \\
&	+ D(\phi) \times (\partial_\mu \partial_\nu \phi) (\partial^\mu\phi) (\partial^\nu\phi)
	+ E(\phi) \times (\partial_\mu \partial_\nu\phi) (\partial^\mu\partial^\nu\phi)
\nonumber \\
&	+ 8 U^{(5)} (\partial_\mu\partial_\nu\partial_\rho \phi) (\partial^\mu\phi) 
		(\partial^\nu\phi) (\partial^\rho\phi)
	+ 24 U^{(4)} (\partial_\mu\partial_\nu\partial_\rho\phi) 
		(\partial^\mu\partial^\nu\phi) (\partial^\rho\phi)
\nonumber \\
&	+ 4 U''' (\partial_\mu\partial_\nu\partial_\rho\phi) 
		(\partial^\mu\partial^\nu\partial^\rho\phi) 
	\biggr\} = 0 \, .
\end{align}
The functions $A$-$E$ are composed out of various derivatives of the potential 
function $U$, and are not important for this discussion. The important thing to 
notice is that the last three terms in the brackets are of third order in derivatives,
and that there is no way of reducing their (time) derivative order while maintaining 
Lorentz covariance. In the first and second order derivative reductions whenever
we needed to use the lower order equation in the higher order part
we were always encountering second time derivatives in the combination
with the Laplacian so that they form the d'Alembertian $\square$. Here this is no
longer the case, and if we want to reduce third derivatives above we need to 
pick out second time derivatives individually, which will break manifest covariance.
This example teaches us that reducing the derivative order at the level of the 
equations of motion is not always compatible with the symmetries of the
system.

Note also that a field redefinition including higher derivatives,
\begin{equation}
\phi \rightarrow \phi - 4 \epsilon^3\alpha _3 U''' \bigl[ 
	(\partial_\mu\partial_\nu \phi) (\partial^\mu \partial^\nu \phi)
	+2 U^{(4)} (\partial_\mu\partial_\nu \phi)
		(\partial^\mu\phi) (\partial^\nu\phi) \bigr] \, ,
\end{equation}
precisely removes the third derivatives from the equation of motion
(and modifies functions $A$-$E$), while preserving covariance. There seems to be
no way around making field redefinitions when perturbatively reducing
the derivative order while maintaining covariance.


\subsubsection{Reduction via perturbative constraints}

Similar maladies afflicting derivative reduction at the level of the
equation of motion exhibited in the previous subsection affect the method of
perturbative constraints. It will be shown here that field redefinitions 
are unavoidable.

The starting point is again the extended action, with time derivatives promoted to
independent fields,
\begin{equation}
\dot{\phi} \rightarrow V_1\, ,
\quad
- \square\phi = \dot{V}_1 - \nabla^2\phi \rightarrow V_2 \, ,
\quad
-\partial_0 \square\phi = \dot{V}_2 \rightarrow V_3 \, ,
\quad
\square\square\phi = \dot{V}_3 - \nabla^2V_2 \rightarrow V_4 \, ,
\end{equation}
and Lagrange multipliers $\pi$, $\pi_1$, $\pi_2$, and $\pi_3$ introduced to 
ensure on-shell equivalence,
\begin{align}
\mathcal{S} ={}& \int\! d^{D\!}x \, \biggl\{
	\frac{1}{2} V_1^2 - \frac{1}{2} (\nabla\phi)^2 - U + \epsilon^3\alpha_3 V_4^2
	+ \pi (\dot{\phi} - V_1) + \pi_1 (\dot{V}_1 - \nabla^2\phi - V_2)
\nonumber \\
&	\hspace{3cm}
	+ \pi_2 (\dot{V}_2 - V_3)
	+ \pi_3 (\dot{V_3} - \nabla^2 V_2 - V_4)
	\biggr\}
\nonumber \\
={}& \int\! d^{D\!}x \, \biggl\{
	\pi\dot{\phi} + \pi_1 \dot{V}_1 + \pi_2 \dot{V}_2 + \pi_3 \dot{V}_3
	- \biggl[
	\pi V_1 + \pi_2 (\nabla^2\phi + V_2)
	+ \pi_2 V_3 
\nonumber \\
&	\hspace{3cm}
	+ \pi_3 (\nabla^2 V_2 +V_4)
	- \frac{1}{2} V_1^2 + \frac{1}{2} (\nabla\phi)^2 + U
	- \epsilon^3 \alpha_3 V_4^2
	\biggr]
	\biggr\} \, .
\label{third order extended action}
\end{align}
We do not bother here with presenting step-by-step how the perturbative 
constraints are imposed
order by order, this should be clear from examples of the first and second order reductions,
and the expressions at third order are rather lengthy while bringing noting new
to the discussion.
Instead we just cite the solutions of the perturbative second-class constraints needed 
to reduce the derivative order, and obtain the lower derivative phase space action,
\begin{align}
\pi_3 \approx{}& - \epsilon^3 \alpha_3 \bigl(  U'' [\pi^2 - (\nabla\phi)^2] - U''U'\bigr) \, ,
\\
\pi_2 \approx{}& - 2 \epsilon^3\alpha_3 \Bigl\{ U^{(4)} \pi [-\pi^2 + (\nabla\phi)^2] 
	- 2 U''' \pi [\nabla^2\phi - U'] 
\nonumber \\
&	\hspace{5cm}	
	+ 2U''' (\nabla\phi) (\nabla\pi)
	+ [U''U']' \pi \Bigr\} \, ,
\\
\pi_1 \approx{}& 2\epsilon^3 \alpha_3 \Bigl\{
	4 U''' (\nabla\pi)^2 + 8 U^{(4)} \pi (\nabla\pi) (\nabla\phi)
	- U^{(5)} [\pi^2-(\nabla\phi)^2]^2
\nonumber \\
&	\hspace{2cm}
	- U^{(4)} \bigl[ 4 \pi^2 \nabla^2\phi + U' (\nabla\phi)^2
		+ 2 (\nabla\phi) \nabla (\nabla\phi)^2 \bigr]
\nonumber \\
&	\hspace{2cm}
	+ U''' \bigl[ - 2 (\nabla^2\phi - U')^2 + 2(\nabla\phi) \nabla[\nabla^2\phi-U']
		- \nabla^2 (\nabla\phi)^2 \bigr]
\nonumber \\
&	\hspace{2cm}
	+ [U''U']''' [\pi^2 - (\nabla\phi)^2]
	+ [5 U^{(4)}U' + 2U'''U''] \pi^2 - [U''U']'U'
	\Bigr\} \, ,
\\
V_4 \approx{}& - U''' [\pi^2 - (\nabla\phi)^2] - U'' U' + \mathcal{O}(\epsilon) \, ,
\\
V_3 \approx{}& - \epsilon^3 U' \, ,
\\
V_2 \approx{}& - U' + \mathcal{O}(\epsilon) \, ,
\\
V_1 \approx{}& \pi + \mathcal{O}(\epsilon) \, .
\end{align}
Promoting these to strong equalities and plugging them in the
extended action~(\ref{third order extended action}) gives the
 phase space reduced action, which takes the form
\begin{equation}
\mathscr{S}_{\rm red} = \int\! d^{D\!}x \, \biggl\{
	8 \epsilon^3 \alpha_3 \bigl[ U''' (\nabla\pi)^2 
		+ 2 U^{(4)} \pi (\nabla\pi) (\nabla\phi) \bigr] \dot{\pi} +
	\bigl( \pi + \epsilon^3\alpha_3 \Pi[\phi,\pi] \bigr) \dot{\phi} - H[\phi,\pi] 
	\biggr\} \, ,
\label{third order phase space}
\end{equation}
where $\Pi$ is a function of $\phi$, $\pi$, and their spatial derivatives,
and so is the Hamiltonian $H$. The method of derivative reduction via perturbative 
second-class constraints accomplishes exactly what it is designed to do
-- produces a phase space formulation of a lower derivative system capturing 
completely the perturbative solutions of the original system to order 
$\mathcal{O}(\epsilon^4)$. But phase space actions are considerably less convenient 
to work with than the configuration space ones, and it is desirable to construct one here.

The specific form of the phase space action~(\ref{third order phase space}) is
of interest for the discussion here, but rather just the symplectic part. 
The fact that the symplectic part appears  appears in a non-canonical
form is by now not surprising, since we have encountered similar situations
at lower orders in derivative reduction. The novel feature here is that 
the symplectic part {\it cannot} be canonicalized by only redefining the 
conjugate momentum field. The part containing the function $\Pi$ can easily
be removed by
\begin{equation}
\pi \rightarrow \pi - \epsilon^3 \alpha_3 \Pi \, ,
\end{equation}
but the part of~(\ref{third order phase space}) containing $\dot{\pi}$, 
cannot be removed this way. There is no way to rewrite it
so that the time derivative is only on the scalar field $\phi$. This is the behaviour
mentioned in Sec~\ref{sec: Formalism} that appears in systems with multiple
degrees of freedom. A field has infinitely many {\it local} degrees of freedom,
and their coupling is expressed through gradients. It is precisely gradients
acting on the conjugate momentum field that are preventing the canonicalization
of the action.

If we insist on not redefining the scalar field itself, this also means there is {\it no}
configuration space formulation of the action~(\ref{third order phase space})
solely in terms of the scalar field $\phi$. There perhaps
is a Lagrangian description with additional constraints, but they are likely 
to require explicit perturbative treatment, which defeats the purpose of
this construction.

The action~(\ref{third order phase space}) can, of course, be canonicalized 
by the scalar field redefinition,
\begin{equation}
\phi \rightarrow \phi + 8\epsilon^3\alpha_3 
	\bigl[ U''' (\nabla\phi)^2 + 2 U^{(4)} \pi (\nabla\pi)(\nabla\phi) \bigr] \, .
\end{equation}
After performing this, we will surely be able to find a configuration space action.
But as soon as we start making such fields redefinitions in the phase space action,
we do not really know (unless we investigate it in detail) whether we are preserving
Lorentz covariance or not, and things become rather complicated.

The conclusion here is that field redefinitions are unavoidable when looking for
a lower derivative configuration space action formulation of the systems where 
higher derivatives enter 
perturbatively. This being the case, we might as well consider field redefinitions
of the configuration space action~(\ref{third order HD action}) right from the 
beginning.


\subsubsection{Reduction via field redefinition}

The field redefinition method when employed at the first and second order yielded
the equivalent results as reducing derivative order in the equation of motion, 
and as the perturbative constraints method. Moreover, it was somewhat easier
and faster to implement. At the third order the difference in the practicality of the
methods becomes huge. In any case we need to employ field redefinitions, but
doing it directly on the configuration space action is now tremendously faster
and more straightforward. With comparatively much less work one can see that the 
scalar field redefinition
\begin{equation}
\phi \rightarrow \phi - \epsilon^3 \alpha^3 \square\square \bigl[ \square\phi - U' \bigr]
	- \epsilon^3 \alpha_3 (U'')^2 (\square\phi) \, ,
\end{equation}
transforms the action~(\ref{third order HD action}) into
\begin{align}
S \rightarrow{}& \int\! d^{D\!}x \, \biggl\{
	- \frac{1}{2} \Bigl( 1 + 2\epsilon^3 \alpha_3 [(U'')^2U']' \Bigr)
		(\partial_\mu\phi) (\partial^\mu\phi) - U
\nonumber \\
&	\hspace{2cm}
	+ \epsilon^3\alpha_3 (U'')^2 (\partial_\mu\phi) (\partial^\mu\phi) 
		(\partial_\nu\phi) (\partial^\nu\phi)
	+ 2\epsilon^3\alpha_3 U''' U'' (\partial_\mu\phi) (\partial^\mu\phi) (\square\phi)
	\biggr\} \, ,
\end{align}
removing all the genuine higher derivative terms, leaving only terms propagating 
three degrees of freedom.


\subsection{Derivative reduction at third order -- Part II}

As a last example of this section we examine the leading
order action~(\ref{leading order action}) with a single higher
order correction~(\ref{Ss(3)7}) from the third order EFT expansion,
\begin{equation}
S = \int\! d^{D\!}x \, \biggl\{
	- \frac{1}{2} (\partial_\mu\phi) (\partial^\mu\phi) - U
	+ \epsilon^3 \alpha_3 (\partial_\mu\phi) (\partial^\mu \phi)
	(\partial_\nu \partial_\rho\phi) (\partial^{\nu}\partial^\rho\phi) 
	\biggr\} \, .
\label{third order HD action II}
\end{equation}
The conclusions of the preceding subsection were that field redefinitions 
are generally inevitable in derivative reduction methods, therefore we 
consider only field redefinition here. 

The higher order correction term in~(\ref{third order HD action II}) is indeed 
a genuine higher derivative term. It is not difficult to recognize that
no Lorentz covariant field redefinition will be able to remove it
(though non-covariant redefinitions would allow for this~\cite{Crisostomi:2017aim}).
At this point we draw the attention to the difference between genuine and 
non-genuine higher derivative terms, former carrying extra degrees of freedom,
compared to the latter that do not but still contain higher (spatial) derivatives.
One such non-genuine higher derivative term is
\begin{equation}
\int\! d^{D\!}x \, (\partial_\mu \phi)(\partial^\nu\phi) \bigl[
	(\square\phi) (\square\phi) 
	- (\partial_\mu\partial_\nu \phi) (\partial^\mu\partial^\nu \phi)
	\bigr] \, ,
\end{equation}
which is a specific flat space specialization of the fourth Horndeski 
action~\cite{Horndeski:1974wa}. We can rewrite the 
action~(\ref{third order HD action II}) in terms of it,
\begin{align}
S ={}& \int\! d^{D\!}x \, \biggl\{
	- \frac{1}{2} (\partial_\mu\phi) (\partial^\mu\phi) - U
	- \epsilon^3 \alpha_3 (\partial_\mu \phi)(\partial^\nu\phi) \bigl[
	(\square\phi) (\square\phi) 
	- (\partial_\mu\partial_\nu \phi) (\partial^\mu\partial^\nu \phi)
	\bigr]
\nonumber \\
&	\hspace{2cm}
	+ \epsilon^3 \alpha_3 (\partial_\mu \phi)(\partial^\nu\phi) (\square\phi) (\square\phi) 
	\biggr\} \, .
\end{align}
Now the only genuine higher derivative term in the action above can
be taken the term in the bottom row, and that one is easily removed by
the following field redefinition,
\begin{equation}
\phi \rightarrow \phi 
	- \epsilon^3 \alpha_3 (\partial_\mu\phi) (\partial^\mu \phi) (\square\phi) \, ,
\end{equation}
transforming the action to
\begin{align}
S\rightarrow{}& \int\! d^{D\!}x \, \biggl\{
	- \frac{1}{2} (\partial_\mu\phi) (\partial^\mu\phi) - U
	- \epsilon^3 \alpha_3 (\partial_\mu \phi)(\partial^\nu\phi) \bigl[
	(\square\phi) (\square\phi) 
	- (\partial_\mu\partial_\nu \phi) (\partial^\mu\partial^\nu \phi)
	\bigr]
\nonumber \\
&	\hspace{2cm}
	+ \epsilon^3 \alpha_3 U' (\partial_\mu \phi)(\partial^\nu\phi) (\square\phi) 
	\biggr\} \, ,
\end{align}
where now the term in the bottom row is not genuinely higher derivative,
and the whole action propagates no extra degrees of freedom. This field
transformation teaches us that sometimes seemingly we do not need to
eliminate a term by field transformation, but produce it in order to
eliminate the genuine higher derivative behavior.


\subsection{Summary of the single scalar case study}

Studying different examples of of perturbative derivative reductions
in single scalar field systems we have learned several lessons summarized
by the following points:
\begin{itemize}
\item
Reducing derivative order at the level of equations of motion is ambiguous.
Many different equations capture perturbatively expandable solutions to
a given order, and only specific ones have an action formulation. 
Those are the ones that should 
be used when attempting to solve the truncated lower 
order equation exactly.

\item
It is not always possible to reduce the derivative order at the  
equations of motion level completely, and at the same time retain 
Lorentz covariance. Field redefinitions are needed to complete the 
task in general. This point was proved in~\cite{Crisostomi:2017aim}
for the case of multiple interacting scalar fields, and second derivatives
in the action.

\item
Derivative reduction by perturbative second-class constraints 
always accomplishes what it is meant to do -- remove spurious degrees of freedom
and deliver a lower derivative (or non-genuine higher derivative) 
phase space action. In general there is no
equivalent configuration space action corresponding to this phase space
reduced one, unless we allow for additional non-covariant field redefinitions.

\item
Field redefinition method successfully eliminates genuine higher derivative terms
from the configuration space action to a desired order, and is consistent
with the first two methods. 

\item
Genuine higher derivative corrections are mapped to lower order or
non-genuine higher derivative corrections. These correctly capture the
EFT corrections. Still, we need to know which higher derivative terms
are not genuine, the reduction method does not tell us this.

\item Removal of genuine higher derivative EFT corrections by field redefinition
is generally not unique (but then again neither is any action where field redefinitions
are allowed).

\end{itemize}
In conclusion, the field redefinition method is the preferred method 
for reducing genuine higher order EFT corrections.


\section{Derivative reduction in scalar-tensor theory}
\label{sec: ST}

In this section we apply the lessons learned in the single scalar case study
of the preceding section, and employ the field redefinition method to the
higher order EFT corrections of a scalar-tensor theory.
Scalar theories are particularly suited for our purposes, since many of the non-genuine
higher derivative terms have been worked out and classified, first
within Horndeski theory~\cite{Horndeski:1974wa}, and later beyond 
Horndeski~\cite{Gleyzes:2014dya,Gleyzes:2014qga}, and lately 
within the more general DHOST 
program~\cite{Langlois:2015cwa,Crisostomi:2016czh,BenAchour:2016fzp}. This actually makes our job much
easier, since we can concentrate only on removing the genuine higher derivative
corrections.

Here we also work in $D$ spacetime dimensions, and use the following conventions:
metric field is of positive signature, the Christoffel symbol is
$\Gamma^{\alpha}_\mu\nu = \frac{1}{2} g^{\alpha\beta} (\partial_\mu g_{\nu\beta}
+ \partial_\nu g_{\mu\beta} - \partial_\beta g_{\mu\nu})$, the Riemann tensor
${R^\rho}_{\mu\sigma\nu} = \partial_\sigma \Gamma^\rho_{\mu\nu} - \dots$,
the Ricci tensor $R_{\mu\nu}={R^\rho}_{\mu\rho\nu}$, the Ricci scalar
$R={R^\mu}_\mu$, and $\nabla_\mu$ is the covariant derivative.

The leading order action in our scalar-tensor theory is assumed to be
\begin{equation}
S^{(0)} = \int\! d^{D\!}x \, \sqrt{-g} \, \biggl\{
	\frac{1}{\kappa^2} f(\phi) R
	- \frac{1}{2} z(\phi) g^{\mu\nu} (\nabla_\mu\phi) (\nabla_\nu\phi)
	- u(\phi)
	\biggr\} \, ,
\qquad
\kappa = \sqrt{16\pi G_N} \, ,
\end{equation}
and it propagates one scalar and two tensor degrees of freedom in $D\!=\!4$.

The EFT expansion of corrections to this action is somewhat awkward with the
metric dimensionless and the scalar field of mass dimension one. It is far more
convenient to define a dimensionless scalar,
\begin{equation}
\varphi = \kappa \phi \, ,
\end{equation}
so that the leading order action reads
\begin{equation}
S^{(0)} = \frac{1}{\kappa^2} \int\! d^{D\!}x \, \sqrt{-g} \, \biggl\{
	F(\varphi) R 
	- \frac{1}{2} Z(\varphi) g^{\mu\nu} (\nabla_\nu \varphi) (\nabla_\nu \varphi)
	- U(\varphi)
	\biggr\}
\label{leading action}
\end{equation}
with the identifications $F(\varphi) = f(\phi)$, $Z(\varphi) = z(\phi)$, 
$U(\varphi) = u(\phi)$.
Now the EFT expansion is naturally organized as an expansion in total number
of derivatives. The leading
order action contains terms with zero and two derivatives. The first order correction
terms then contain four derivatives, and are one $\kappa^2$ order higher then the
leading order action,
\begin{equation}
S^{(1)} = \sum_{n=1}^{9} S^{(1)}_n \, ,
\end{equation}
where all the independent terms containing four derivatives are
\begin{align}
S^{(1)}_1 ={}& \int\! d^{D\!}x \, \sqrt{-g} \, f_1(\varphi) \,
	 (\nabla_\mu\varphi) (\nabla^\mu\varphi) (\nabla_\nu\varphi) (\nabla^\nu\varphi) \, ,
\label{S(1)1}
\\
S^{(1)}_2 ={}& \int\! d^{D\!}x \, \sqrt{-g} \, f_2(\varphi) \, 
	(\nabla_\mu \varphi) (\nabla^\mu \varphi) (\square\varphi) \, ,
\label{S(1)2}
\\
S^{(1)}_3 ={}& \int\! d^{D\!}x \, \sqrt{-g} \, f_3(\varphi) \, 
	(\square\varphi) (\square\varphi) \, ,
\label{S(1)3}
\\
S^{(1)}_4 ={}& \int\! d^{D\!}x \, \sqrt{-g} \,  f_4(\varphi) \, 
	(\nabla_\mu\varphi) (\nabla^\mu\varphi) R \, ,
\label{S(1)4}
\\
S^{(1)}_5 ={}& \int\! d^{D\!}x \, \sqrt{-g} \,  f_5(\varphi) \, 
	(\nabla_\mu\varphi) (\nabla_\nu\varphi) R^{\mu\nu}  \, ,
\label{S(1)5}
\\
S^{(1)}_6 ={}& \int\! d^{D\!}x \, \sqrt{-g} \,  f_6(\varphi) \, 
	(\square \varphi) R \, ,
\label{S(1)6}
\\
S^{(1)}_7 ={}& \int\! d^{D\!}x \, \sqrt{-g} \,  f_7(\varphi) \, R^2 \, ,
\label{S(1)7}
\\
S^{(1)}_8={}& \int\! d^{D\!}x \, \sqrt{-g} \,  f_8(\varphi) \, R_{\mu\nu} R^{\mu\nu} \, ,
\label{S(1)8}
\\
S^{(1)}_9={}& \int\! d^{D\!}x \, \sqrt{-g} \,  f_9(\varphi) \, 
	R_{\mu\nu\rho\sigma} R^{\mu\nu\rho\sigma} \, ,
\label{S(1)9}
\end{align}
and $f_1$-$f_9$ are arbitrary functions of the scalar field. All the other
scalar-tensor actions containing four derivatives that one can write down can
be expressed in terms if the nine above my making use of partial integrations,
commutators of covariant derivatives, and the Bianchi identity.

Not all of the terms~(\ref{S(1)1}-\ref{S(1)9}) 
are genuinely higher-derivative ones, in the sense that they do not introduce
additional degrees of freedom compared to the three present in the leading
order action. 
Luckily, the most general scalar-tensor action containing four derivatives 
and propagating only three degrees of freedom is known. 
Most of it is captured
by the Horndeski action~\cite{Horndeski:1974wa} 
specialized to four derivatives\footnote{
In the notation standard in the Horndeski theory literature we have
(using conventions $X= (\partial_\mu\varphi) (\partial^\mu\varphi)$)
\vskip-0.8cm
\begin{equation*}
G_2(\varphi, X) = g_2(\varphi) X^2 \, ,
\qquad 
G_3(\varphi, X) = g_3(\varphi) X \, ,
\qquad
G_4(\varphi, X) = g_4(\varphi) X \, ,
\qquad
G_5(\varphi, X) = 0 \, .
\end{equation*}
},
\begin{align}
S^{\rm H}_2 ={}& \int\! d^{D\!}x \, \sqrt{-g} \, g_2(\varphi) \, 
	(\nabla_\mu \varphi) (\nabla^\mu \varphi) 
	(\nabla_\nu \varphi) (\nabla^\mu \varphi) \, ,
\label{SH2}
\\
S^{\rm H}_3 ={}& \int\! d^{D\!}x \, \sqrt{-g} \, g_3(\varphi) \,
	(\nabla_\mu \varphi) (\nabla^\mu \varphi) (\square\varphi) \, ,
\label{SH3}
\\
S^{\rm H}_4 ={}& \int\! d^{D\!}x \, \sqrt{-g} \, g_4(\varphi) \,
	\Bigl\{ (\nabla_\mu\varphi) (\nabla^\mu\varphi) R 
		- 2 \bigl[ (\square\varphi) (\square\varphi) 
		- (\nabla_\mu\nabla_\nu \varphi) (\nabla^\mu \nabla^\nu \varphi) \bigr]
	\Bigr\} \, .
\label{SH4}
\end{align}
The last of these actions can be rearranged into a more convenient form
\begin{align}
S^{\rm H}_4 ={}&
 \int\! d^{D\!}x \, \sqrt{-g} \, \biggl\{
	- 2 g_4(\varphi) \, G^{\mu\nu} (\nabla_\mu \varphi) (\nabla_\nu \varphi)
	+ 3 g_4'(\varphi) \, (\nabla_\mu\varphi) 
		(\nabla^\mu \varphi) (\square\varphi)
\nonumber \\
&	\hspace{4cm}
	+ g_4''(\varphi)\, (\nabla_\mu\varphi) (\nabla^\mu \varphi)
		(\nabla_\nu\varphi) (\nabla^\nu \varphi)
	\biggr\} \, ,
\label{StH4}
\end{align}
so instead of it we will use a more convenient one
\begin{equation}
\widetilde{S}^{\rm H}_4 = \int\! d^{D\!}x \, \sqrt{-g} \,
	\widetilde{g}_4(\varphi) \, G^{\mu\nu} 
	(\nabla_\mu\varphi) (\nabla_\nu\varphi) \, .
\end{equation}
There is one more remaining action with four derivatives propagating only
three degrees of
freedom, and it is known from Lanczos-Lovelock 
gravity~\cite{Lovelock:1971yv,Lanczos:1938sf},
and more generally from scalar-Gauss-Bonnet models~\cite{Nojiri:2005vv},
\begin{equation}
S^{\rm SGB} = \int\! d^{D\!}x \, \sqrt{-g} \, 
	b(\varphi) \bigl[ R^2 - 4R_{\mu\nu} R^{\mu\nu} 
		+ R_{\mu\nu\rho\sigma} R^{\mu\nu\rho\sigma} \bigr]
\label{SGB}
\end{equation}
The actions~(\ref{SH2}), (\ref{SH3}), (\ref{StH4}), and~(\ref{SGB}) comprise the most 
general four-derivative action propagating one scalar and two tensor degrees
of freedom.

Out of nine independent four-derivative actions~(\ref{S(1)1}-\ref{S(1)9})
four do not need to be considered as higher derivative since we can instead
write them as combinations forming actions~(\ref{SH2}), (\ref{SH3}), (\ref{StH4}), 
and~(\ref{SGB}) which are not genuinely higher derivative. They can be added
to the leading order scalar-tensor action~(\ref{leading action}) 
as genuine EFT corrections
not introducing spurious degrees of freedom.
These considerations reduce the number of genuinely higher derivative terms
we need to worry about to only five, which we relabel for convenience,
\begin{align}
S^{\rm HD}_1 ={}& S_8^{(1)} = \int\! d^{D\!}x \, \sqrt{-g} \,
	h_1(\varphi) \, R_{\mu\nu} R^{\mu\nu} \, ,
\label{SHD1}
\\
S^{\rm HD}_2 ={}& S_7^{(1)} = \int\! d^{D\!}x \, \sqrt{-g} \,
	h_2(\varphi) \, R^2 \, ,
\label{SHD2}
\\
S^{\rm HD}_3 ={}& S_6^{(1)} = \int\! d^{D\!}x \, \sqrt{-g} \,
	h_3(\varphi) \, (\square\varphi)  R \, ,
\label{SHD2}
\\
S^{\rm HD}_4 ={}& S_5^{(1)} = \int\! d^{D\!}x \, \sqrt{-g} \,
	h_4(\varphi) \, (\nabla_\mu\varphi) (\nabla^\mu \varphi) R \, ,
\label{SHD4}
\\
S^{\rm HD}_5 ={}& S_3^{(1)} = \int\! d^{D\!}x \, \sqrt{-g} \,
	h_5(\varphi) \, (\square\varphi) (\square\varphi) \, .
\label{SHD5}
\end{align}
%
%
%

We proceed to eliminate the five higher derivative 
corrections~(\ref{SHD1}-\ref{SHD5}) to the action
by field redefinitions of the dimensionless scalar and the metric,
\begin{equation}
\varphi \rightarrow \varphi + \kappa^2 \delta\varphi \, ,
\qquad
g_{\mu\nu} \rightarrow g_{\mu\nu} + \kappa^2 \delta g_{\mu\nu} \, .
\label{field redef}
\end{equation}
The leading order action~(\ref{leading action}) transforms under 
these field redefinitions as
\begin{equation}
S^{(0)} \rightarrow S^{(0)} + \kappa^2 [\delta_\varphi S^{(0)}] 
	+ \kappa^2 [\delta_g S^{(0)} ] \, ,
\end{equation}
where the first subleading term on the right hand side descends form the scalar
field redefinition,
\begin{equation}
\kappa^2 [\delta_\varphi S^{(0)}] = \int\! d^{D\!}x \, \sqrt{-g} \, \biggl\{
	F' R  + Z(\square\varphi) 
	- \frac{1}{2} Z' (\nabla_\mu\varphi) (\nabla^\mu\varphi) - U' 
	\biggr\} \delta\varphi \, ,
\label{action phi transformation}
\end{equation}
and the second one descends from the metric field redefinition,
\begin{align}
\MoveEqLeft[3]
\kappa^2 [\delta_g S^{(0)}] = \int\! d^{D\!}x \, \sqrt{-g} \, \biggl\{
	- F G_{\mu\nu} 
	+ F' \bigl[ (\nabla_\mu\nabla_\nu \varphi) - (\square\varphi) g_{\mu\nu} \bigr]
\nonumber \\
&
	+ \frac{1}{2} [Z+2F''] (\nabla_\mu \varphi) (\nabla_\nu \varphi)
	- \frac{1}{4} [Z + 4 F''] (\nabla_\rho \varphi) (\nabla^\rho \varphi) g_{\mu\nu}
	- \frac{1}{2}U g_{\mu\nu}
	\biggr\} \delta g^{\mu\nu} \, .
\label{action g transformation}
\end{align}

In order to accomplish the task of removing higher derivatives from the first
order corrections~(\ref{SHD1}-\ref{SHD5}) while preserving general covariance,
quantities $\delta\varphi$ and $\delta g_{\mu\nu}$ have 
to contain two derivatives and have to be covariant themselves.
Their general form is thus
\begin{align}
\delta\varphi ={}& C_1\times (\square\varphi)
	+ C_2 \times (\nabla_\mu \varphi) (\nabla^\mu \varphi) 
	+ C_3 \times R \,
\label{scalar field redef}
\\
\delta g_{\mu\nu} ={}& C_4 \times g_{\mu\nu} R
	+ C_5 \times  R_{\mu\nu}
	+ C_6 \times g_{\mu\nu} (\square \varphi)
\nonumber \\
&	\hspace{1cm}
	+ C_7 \times  (\nabla_\mu \! \nabla_\nu \varphi)
	+ C_8 \times g_{\mu\nu} (\nabla_\rho\varphi) (\nabla^\rho\varphi)
	+ C_9 \times (\nabla_\mu\varphi) (\nabla_\nu\varphi) \, .
\label{metric field redef}
\end{align}
where the coefficients $C_1$-$C_9$ are functions of the scalar field, $C_i \!=\! C_i(\varphi)$.
We plug these into the expressions for the transformation of leading order 
action,~(\ref{action phi transformation}) and~(\ref{action g transformation}),
and after performing some partial integrations and making use of the Bianchi
 identity,  the transformed leading order action is put into the form
\begin{equation}
S^{(0)} \rightarrow S^{(0)} + \int\! d^{D\!}x\, \sqrt{-g} \, \biggl\{
	\sum_A B_A(\varphi) \times \mathcal{R}_A[g_{\mu\nu}, \nabla_\mu\varphi]
	\biggr\} \, ,
\label{full action transform}
\end{equation}
where all the terms $\mathcal{R}_A$ dependent on the metric 
and the covariant derivatives of the scalar field, 
and the scalar dependent functions $B_A$ are
given in Table~\ref{transformation coefficients}.

\begin{table}[h!]
\setlength{\tabcolsep}{8pt}
\def\arraystretch{1.7}
\centering
\begin{tabular}{|c||c|} 
\hline
$\mathcal{R}_A$ & $B_A(\varphi)$ \\
\hline\hline
$R_{\mu\nu} R^{\mu\nu}$ & $- C_5 F$
\\ \hline
$R^2$ & $C_3 F' + \bigl( \frac{D-2}{2} \bigr) C_4 F + \frac{1}{2} C_5 F$
\\ \hline
$R \, (\square\varphi)$ & 
	$C_1 F' + C_3 Z - (D\!-\!1) C_4 F' + \bigl( \frac{D-2}{2} \bigr) C_6 F$
\\ \hline
$R \, (\nabla_\mu \varphi) (\nabla^\mu \varphi)$ &	
	$C_2 F' - \frac{1}{2} C_3 Z' - \frac{1}{4} C_4 \bigl[ (D\!-\!2)Z + 4(D\!-\!1) F'' \bigr]
	+ \frac{1}{2} C_5' F' $
\\
&	$- \frac{1}{2} C_7 F' + \bigl( \frac{D-2}{2} \bigr) C_8 F$
\\ \hline
$(\square\varphi) (\square\varphi)$ & $C_1 Z - (D\!-\!1) C_6 F' $
\\ \hline
$G^{\mu\nu} (\nabla_\mu\varphi) (\nabla_\nu\varphi)$ & 
	$\frac{1}{2} \bigl[ C_5 Z - 2 C_5' F' \bigr] + C_7' F
	- C_9 F$
\\ \hline
$(\nabla_\mu\varphi) (\nabla^\mu\varphi) (\square\varphi)$ & 
	$- \frac{1}{2} C_1 Z' + C_2 Z - \frac{1}{4} C_6 \bigl[ (D\!-\!2)Z + 4(D\!-\!1) F'' \bigr]$
\\
&	$+ \frac{1}{2} \bigl[ 3 C_7' F' - C_7 Z \bigr] - (D\!-\!1)C_8 F' - \frac{3}{2} C_9 F'$
\\ \hline
$(\nabla_\mu\varphi) (\nabla^\mu\varphi) (\nabla_\nu\varphi) (\nabla^\nu\varphi)$ & 
	$- \frac{1}{2} C_2Z' + \frac{1}{4} \bigl\{ 2 [C_7' F']' - [C_7 Z]' \bigr\}
	- \frac{1}{4} \bigl[ (D\!-\!2) Z + 4(D\!-\!1) F'' \bigr]$
\\
&	$+ \frac{1}{4} \bigl\{ C_9 Z - 2[C_9F']' \bigr\}$
\\ \hline
$R$ & $ -C_3 U' - \frac{D}{2} C_4 U - \frac{1}{2} C_5 U$
\\ \hline
$(\nabla_\mu\varphi) (\nabla^\mu\varphi)$ & 
	$[C_1U']' - C_2 U' + \frac{D}{2} [C_6U]' + \frac{1}{2} [C_7U]'
	- \frac{D}{2} C_8 U - \frac{1}{2} C_9 U$
\\ \hline
\end{tabular}
\caption{All the terms appearing in the transformation of the leading order
action~(\ref{full action transform}) due to field 
redefinitions~(\ref{field redef},\ref{scalar field redef},\ref{metric field redef}).
\label{transformation coefficients}}
\end{table}

We need to choose the nine coefficient functions $C_i$'s so as to
absorb five distinct terms~(\ref{SHD1}-\ref{SHD5}), and the way to do 
this is not unique. In particular, the first five rows of 
Table~\ref{transformation coefficients} need to absorb the genuine higher
derivative terms.
For the argument that we intend to make here it suffices to demonstrate that one
particular solution exists,
\begin{align}
& C_2 = C_6 = C_7 = C_9 = 0 \, ,
\qquad
C_1 = - \frac{h_5}{Z} \, ,
\qquad
C_5 = \frac{h_1}{F} \, ,
\\
& C_4 = \frac{- Z h_1 + 2 Z h_2 - 2 F' h_3 + \frac{2(F')^2}{Z} h_5}
	{(D\!-\!2)ZF + 2(D\!-\!1) (F')^2} \, ,
\qquad
 C_3 = \frac{1}{Z} \biggl\{
	- h_3 + \frac{F'}{Z} h_5 + (D\!-\!1) F' C_4
	\biggr\} \, ,
\\
& C_8 = \frac{-2}{(D\!-\!2)F} \biggl\{ h_4 + \frac{F'}{F}h_1' - \frac{(F')^2}{2F^2} h_1
	- \frac{Z'}{2} C_3 - \frac{1}{4} \bigl[ (D\!-\!2)Z + 4(D\!-\!1)F'' \bigr] C_4 \biggr\} \, .
\end{align}

Therefore, the first order correction in the EFT expansion of the scalar-tensor
theory is captured correctly by the most general scalar-tensor theory with
at most four derivatives that contains only three (in $D\!=\!4$) degrees of freedom,
\begin{align}
S ={}& \int\! d^{D\!}x \, \sqrt{-g} \Biggl\{  
	\frac{1}{\kappa^2} \biggl[ F_1 \, R
	- \frac{1}{2} F_2
		(\nabla_\mu \varphi) (\nabla^\mu \varphi)
	-  F_3 \biggr]
	+ F_4 \, (\nabla_\mu\varphi) (\nabla^\mu\varphi) 
		(\nabla_\nu\varphi) (\nabla^\nu\varphi)
\nonumber \\
&
	+ F_5 \, (\nabla_\mu \varphi) (\nabla^\mu \varphi) (\square \varphi)
	+ F_6 \, G^{\mu\nu} (\nabla_\mu \varphi) (\nabla_\nu \varphi)
	+ F_7 \, \bigl( R^2 - R_{\mu\nu} R^{\mu\nu}
		- R_{\mu\nu\rho\sigma} R^{\mu\nu\rho\sigma} \bigr)
	\Biggr\} \, ,
\end{align}
where $F_1$-$F_7$ are some functions of the scalar field, $F_i \!=\! F_i(\varphi)$
that in principle have an expansion in $\kappa^2$. This action is a subset of
Horndeski actions.


\section{Conclusions}

The aim of this paper was to examine the nature of higher derivative corrections 
in the effective field theory expansions. The point of view that we take on them
is that they arise as a peculiarity of the approximation scheme 
of local derivative expansion employed in the EFT expansion. The higher derivative 
corrections cannot be treated on the same footing as the lower derivative
leading order action describing the system. Rather, their effect should be taken
into account in the same spirit in which they arise -- perturbatively. That implies
capturing their effects with terms that do not introduce extra degrees of freedom.
Three methods of derivative reduction of such higher derivative terms were
examined and compared on several examples of single scalar field systems,
at different order in perturbation parameter.

The examined methods are (i) derivative reduction at the equations of motion
level, (ii) derivative reduction via perturbative constraints, and (iii) derivative reduction
via field redefinitions. The lessons learned are summarized in the bullet points at 
the end of Sec.~\ref{sec: Scalar}. The main conclusion is that the field redefinition
method is the most practical one, compatible with global (Lorentz) symmetries,
yielding as a result a configuration space action with no genuine higher derivative
corrections to a given order. 

Few words are in order on why field redefinition is a legitimate method to use.
When employed to eliminate higher derivative terms to certain order
of interest, field redefinition necessarily involves derivatives itself which
is usually not admissible. It is often 
pragmatically argued that in EFTs this is admissible due to 
pushing the problem of higher 
derivatives beyond the perturbative order of interest. Even though this
rationale is strictly not wrong, a stronger, but more subtle argument
can be made. In general, nonlocal field redefinitions are admissible if
they do not change the number of degrees of freedom in the action.
We can imagine such nonlocal field redefinitions that have a local derivative
expansion analogous to the derivative expansion of the nonlocal correction
to the action defining the EFT expansion. In particular, we can imagine a
special nonlocal field redefinition whose local expansion is responsible for
cancelling the genuine higher derivative terms from the EFT expansion
of the action. In this sense the presence of higher derivatives in the field
redefinition is not seen as fundamental, but just as a consequence of the 
approximation scheme. This is the reasoning behind employing field
redefinitions to remove spurious degrees of freedom from the EFT expansion.

Genuine higher derivatives in EFTs are thus seen as artefacts of the approximation
scheme, not as essential features. Hence, there is no reason to look for a UV
principle that would forbid their appearance in the low energy EFTs, since the. Rather,
we ought to write EFT expansions with an additional requirement -- no terms in
the expansion should be genuine higher derivative terms, {\it i.e.} we should
not write the terms introducing new degrees of freedom. This would remedy
the error we are making by using implicitly the local derivative expansion as the
definition of the EFT. We can make a claim that the {\it EFTs should consist of all the 
possible corrections respecting assumed symmetries, and not introducing new
degrees of freedom}. This work does not provide a general proof of this statement,
but it certainly seems compelling in the light of examples presented here.

Observation about EFTs has made here bear relevance for the scalar-tensor 
theories often studied today. The most general scalar-tensor theories
with no genuine higher derivative terms~\cite{Langlois:2015skt} 
go under the acronym DHOST (Degenerate-Higher-Order-Scalar-Tensor) .
Here we make a claim that, together with the most general such theories
involving curvature tensors as well, they should capture completely all the possible
EFT corrections, and can in fact be seen as arising as EFTs.

Here we have shown explicitly that  the first order correction in the 
EFT expansion of the scalar-tensor
theory is captured correctly by the most general scalar-tensor theory with
at most four derivatives that does not contain extra degrees of freedom,
\begin{align}
S ={}& \int\! d^{D\!}x \, \sqrt{-g} \, \Biggl\{  
	\frac{1}{\kappa^2} \biggl[ F_1 \, R
	- \frac{1}{2} F_2
		(\nabla_\mu \varphi) (\nabla^\mu \varphi)
	-  F_3 \biggr]
	+ F_4 \, (\nabla_\mu\varphi) (\nabla^\mu\varphi) 
		(\nabla_\nu\varphi) (\nabla^\nu\varphi)
\nonumber \\
&
	+ F_5 \, (\nabla_\mu \varphi) (\nabla^\mu \varphi) (\square \varphi)
	+ F_6 \, G^{\mu\nu} (\nabla_\mu \varphi) (\nabla_\nu \varphi)
	+ F_7 \, \bigl( R^2 - R_{\mu\nu} R^{\mu\nu}
		- R_{\mu\nu\rho\sigma} R^{\mu\nu\rho\sigma} \bigr)
	\Biggr\} \, ,
\label{final action}
\end{align}
where $F_1$-$F_7$ are some functions of the scalar field, $F_i \!=\! F_i(\varphi)$
that in principle have an expansion in $\kappa^2$. Even though higher derivatives
appear in this action they are not genuine higher derivatives. In
$D\!=\!4$ spacetime dimensions this action propagates only one scalar and two
tensor degrees of freedom. 

It is worth noting that just recently Solomon and Trodden~\cite{Solomon:2017nlh}
 have made many of the points presented in this paper, and have performed similar
 analyses.


It is now possible to address the initial motivation for this study -- the construction
and quantization of gauge-invariant cosmological perturbations to cubic order in small 
diffeomorphism transformations.
This should to be possible with using the action~(\ref{final action}).
The perturbations of the fields are introduced as 
$g_{\mu\nu} = \overline{g}_{\mu\nu} + \kappa \delta g_{\mu\nu}$ and
$\varphi = \overline{\varphi} + \kappa \delta \varphi$, and the background fields
are assumed to satisfy the leading order equation of motion. The 
higher order terms in the action will modify the structure of the gauge-invariant
perturbations at quadratic
and cubic order. Functions $F_1$-$F_7$ should accommodate
the one-loop renormalization of the two-point functions of gauge-invariant
perturbations, by absorbing the divergences into their parameters.
This is consistent with a well known result regarding the 
absence of one-loop divergences in pure gravity~\cite{tHooft:1974toh}.
Namely, the ones proportional to $R^2$ and $R_{\mu\nu} R^{\mu\nu}$
can be absorbed by the metric field redefinition, in the same way as
we have considered it in Sec.~\ref{sec: ST}, and the remaining one can only
be a pure Gauss-Bonnet term, which is a total derivative in $D\!=\!4$.


General proofs of the statements made here about the absence of genuine
higher derivative corrections in EFTs would be very very powerful, as the 
authors of~\cite{Burgess:2014lwa} point out. It should be pointed out that
the difficult part in any such proof at higher orders than ones considered here
is classifying the non-genuine higher derivative corrections which need not
be removed. This cannot be done by simply counting the number of derivatives
appearing in the action, but rather one often has to resort to canonical
formalism to be able to count the number of degrees of freedom 
correctly~\cite{Langlois:2015skt} .


\acknowledgements

The author is indebted to Tomislav Prokopec for useful suggestions.
This work was supported by the grant 2014/14/E/ST9/00152 of the Polish National
Science Centre (NCN).





\end{document}